\documentclass[12pt,preprint]{aastex}

\usepackage{graphicx}
\usepackage{rotating}
\shorttitle{}
\shortauthors{Nesvorn\'y et al.}

\begin{document}
\baselineskip 19.pt

\title{Catalog of Proper Orbits for 1.25 Million Main Belt Asteroids \\ 
and Discovery of 136 New Collisional Families}

\author{David Nesvorn\'y$^1$, Fernando Roig$^2$, David Vokrouhlick\'y$^3$, 
Miroslav Bro\v{z}$^3$}

\affil{(1) Department of Space Studies, Southwest Research Institute, 1050 Walnut St., 
  Suite 300,  Boulder, CO 80302, USA}

\affil{(2) Observat\'orio Nacional, Rua Gal. Jose Cristino 77, Rio de Janeiro, 
RJ 20921-400, Brazil}

\affil{(3) Institute of Astronomy, Charles University, V Hole\v{s}ovi\v{c}k\'ach 2, 
CZ–18000 Prague 8, Czech Republic}

\begin{abstract}
The proper elements of asteroids are obtained from the instantaneous orbital elements 
by removing periodic oscillations produced by gravitational interactions with planets.
They are unchanging in time, at least if chaotic dynamics and non-gravitational forces 
could be ignored, and can therefore be used to identify fragments of major collisions 
(asteroid families) that happened eons ago. Here we present a new catalog of proper 
elements for $1.25\times10^6$ main belt asteroids. We explain the methodology, evaluate 
uncertainties, and discuss how the new catalog can be used to identify asteroid 
families. A systematic search for families yielded 153 cases not reported in
Nesvorn\'y at al. (2015) -- 17 of these cases were identified in various other 
publications, 136 cases are new discoveries. There are now 274 families in the
asteroid belt in total (plus a handful of families in the resonant Hilda population). 
We analyzed several compact families in detail. The new family around the middle belt
asteroid (9332) 1990SB1 (9 members) is the youngest family found so far (estimated
formation only 16-17 kyr ago). 
New families (1217) Maximiliana, (6084) Bascom, (10164) Akusekijima and (70208) 1999RX33
all formed 0.5-2.5 Myr ago. The (2110) Moore-Sitterly family is a close pair of relatively
large bodies, 2110 and 44612, and 15 small members all located sunwards from 2110 and
44612, presumably a consequence of the Yarkovsky drift over the estimated family age
(1.2-1.5 Myr). A systematic characterization of the new asteroid families is left for
future work.  
\end{abstract}

\section{Introduction}

The orbital elements of asteroids change with time due to gravitational perturbations
of planets and other major bodies (e.g. Ceres). The changes can be periodic (i.e.,
oscillatory) or aperiodic (e.g., chaotic diffusion), and can 
happen on different timescales. The proper elements are computed from normal (osculating)
orbital elements by removing the oscillatory terms. They remain roughly constant on 
long time intervals, which is useful for studies of asteroid families (e.g., Zappal\`a et al. 
1990, 1994; Parker et al. 2008; Milani et al. 2014; Nesvorn\'y et al. 2015; 
Novakovi\'c et al. 2022). Sophisticated analytic methods were traditionally employed to this purpose 
(Milani \& Kne\v{z}evi\'c 1994), but the ever increasing computer power now allows the 
proper elements to be computed to a greater precision numerically (\v{S}idlichovsk\'y \&
Nesvorn\'y 1996; Kne\v{z}evi\'c \& Milani 2000, 2019). The proper element catalog updates
at the AstDys node\footnote{\texttt{https://newton.spacedys.com/astdys/}} have been discontinued 
-- they are hosted by B. Novakovi\'c at the Asteroid Families Portal (AFP; Novakovi\'c et 
al. 2022).\footnote{\texttt{http://asteroids.matf.bg.ac.rs/fam/}} 

The three most useful proper elements are: the proper semimajor axis ($a_{\rm p}$), the 
proper eccentricity ($e_{\rm p}$), and the proper inclination ($i_{\rm p}$). They are close 
equivalents to their osculating element counterparts in that they define the average size, 
elongation and tilt of orbits, respectively. Here we use numerical methods (Sect. 2)
to compute $a_{\rm p}$, $e_{\rm p}$ and $i_{\rm p}$ for main-belt asteroids listed 
in the most recent Minor Planet Center (MPC) catalog. The new catalog of proper orbits 
is publicly available.\footnote{At \texttt{https://asteroids.on.br/appeal/}, 
\texttt{www.boulder.swri.edu/\~{}davidn/Proper24/}, 
and the Planetary Data System (PDS) node, \texttt{https://pds.nasa.gov/}} The methodology 
described below is scalable to $\sim 10^7$ bodies and can be used to compute 
proper orbits for the large volume of main-belt asteroids expected to be discovered 
by the Rubin Observatory in the next decade ({see Sect. 5}; Schwamb et al. 2023).

\section{Methods}

\subsection{Proper elements}

The orbital elements of main belt asteroids were downloaded from the MPC catalog on 
February 9, 2024. We selected asteroid orbits with $a>1.6$ au and the perihelion distance 
$q>1.3$ au (to avoid near-Earth asteroids), aphelion distance $Q<5$ au (to avoid 
unstable Jupiter-crossing orbits), and $a<3.8$ au (to avoid Hildas in the 3:2 
resonance with Jupiter).\footnote{A different strategy must be employed to compute 
  the proper elements for Hildas (Bro\v{z} \& Vokrouhlick\'y 2008,
  Bro\v{z} et al. 2011).} This represents 1,261,151 
orbits in total. We do not distinguish between the numbered and unnumbered (single- or 
multi-opposition) bodies, but the information about the quality of the osculating orbits 
(the number of oppositions) is propagated to the final catalog. The osculating 
orbits are given with respect to the J2000 ecliptic reference system and, for the 
vast majority of cases, at the JD 2460200.5 epoch (a small number of orbits given at
different epochs are ignored). The planetary orbits (Mercury to Neptune) were obtained 
for the same epoch from the DE 441 Ephemerides (Park et al. 2021). We used the center of 
mass -- planet plus its satellites -- and the total mass of each system. The gravitational 
effects of Ceres and other massive asteroids were ignored (see Tsirvoulis \& 
Novakovi{\'c} 2016).\footnote{We also performed the analysis with Ceres being included as 
a massive body in the orbital integrations. As the computed proper elements do not show any 
significant differences, here we report the results for the case without the gravitational 
effects of Ceres.} The reference system was rotated to the invariant plane of planets 
(as defined by the total angular momentum of planetary orbits). 

The orbital integrations were performed with the \texttt{Swift} integrator (Levison \& Duncan
1994; code \texttt{swift\_{}rmvs4}), which is an efficient implementation of the Wisdom-Holman map
(Wisdom \& Holman 1991). We used a short time step (1.1 days) and integrated all orbits backward
in time for 10 Myr.\footnote{We tested different timesteps. The one that was selected for the
  main integrations is conservatively short. We checked that the integrated orbit of Mercury is
  stable and does not show any unusual behavior. The semimajor axis of Mercury shows oscillations
  around a fixed value (i.e., no diffusion) and Mercury's eccentricity/inclination evolution
  closely follows expectations. The proper frequencies of Mercury are correctly recovered from the
  integration. We also recalculated the proper elements of the first 1000 asteroids from an integration
  where we halved the time step and found that the proper elements and their errors were practically
  identical to those obtained from the original integration.}
The backward integration is useful to identify any past convergence of angles, which may
indicate the formation time of a young asteroid family (Nesvorn\'y et al. 2002). The symplectic 
corrector was applied to compensate for high-frequency noise terms (Wisdom 2006).  
We adopted the general relativistic correction from Quinn et al. (1991). The integrations 
were split over 12,620 Ivy-bridge cores on the NASA's Pleiades Supercomputer and ran for 60 
wall-clock hours. The orbital elements were saved in double precision every 600 years for the 
total of 16,666 outputs per orbit.\footnote{We modified the original output scheme in 
\texttt{Swift} such that the fixed output cadence is strictly enforced.} This allows us to
resolve frequencies as high as 1000 arcsec/yr. The binary output files 
represent 1.3 Tb of data in total. We did not apply any low-pass filter on output, because our 
previous tests showed that the use of filter did not have an appreciable effect on the 
final product. 

The Frequency Modified Fourier Transform (FMFT; \v{S}idlichovsk\'y \& Nesvorn\'y 1996,
Laskar 1993) was applied to obtain a Fourier decomposition of each signal. We used the complex
variable $x(t) + \iota y(t)$ with $x=e\cos(\varpi)$ and $y=e\sin (\varpi)$ for 
the proper eccentricity, and $x=\sin(i)\cos(\Omega)$ and $y=\sin(i) \sin (\Omega)$ for the 
proper inclination, where $\varpi$ and $\Omega$ are the perihelion and nodal longitudes. 
FMFT was first applied to the planetary orbits to obtain the planetary frequencies $g_j$ and 
$s_j$, governing the perihelion and nodal evolution of planetary orbits, respectively 
(Table 1). We identified the forced terms with these frequencies in the Fourier decomposition 
of each asteroid orbit, and subtracted them from asteroid's $x(t) + \iota y(t)$.

We experimented with different techniques to extract the amplitude of the proper terms from
the remaining signal. It is possible, for example, to identify the proper frequency as the
largest term in the remaining signal and use FMFT to obtain its amplitude. This is the method
recommended in Kne\v{z}evi\'c \& Milani (2000). It works perfectly well in the vast 
majority of cases. For orbits near mean-motion and secular resonances, however, 
which may be affected by orbital chaos, we observed splitting of the proper term into a number 
of Fourier terms with similar frequencies. In these cases, the amplitude of the largest 
proper term usually corresponds to the {\it minimum} of ($e$ or $\sin (i)$) oscillations, 
which is inconvenient because the normal proper elements are desired to be close to the 
{\it mean} value of osculating elements (Appendix A). 

Short Fourier intervals could be used to reduce problems with the proper term splitting, but the 
optimal interval length is unknown a priori. Also, as long intervals should be used to 
define more stable proper elements in most cases, one would be tempted to use the Fourier interval 
that flexibly adjusts from case to case. Unfortunately, according to our tests, it is not obvious 
how to define robust criteria for the variable interval length.

We therefore opted for a more reliable method which simply consists in computing the mean
of $\sqrt{x(t)^2+y(t)^2}$, with the forced terms removed, over a relatively long interval (2.5 or 5 Myr).
We found that the proper elements computed from a longer interval generally have better precision;
the 5-Myr window was adopted for the final catalog.
This defines the proper elements $e_{\rm p}$ and $\sin i_{\rm p}$. Following Kne\v{z}evi\'c \& Milani 
(2000), the proper semimajor axis was computed as the mean value of the osculating semimajor axis
over the same time interval. For asteroids in mean-motion resonances, this means that $a_{\rm p}$ 
falls near the $a$ value of the exact resonance (e.g., Nesvorn\'y et al. 2002). The uncertainties 
of $a_{\rm p}$, $e_{\rm p}$ and $\sin i_{\rm p}$ were obtained as the RMS of the proper elements computed 
from different intervals within the 10-Myr integration time span.\footnote{We tested different choices
  and, for the uncertainties reported in the final catalog, used five equally spaced intervals that
  cover the whole integration range. Using more intervals slows down the calculation but does not
  significantly improve the estimate of uncertainties.}

\subsection{Asteroid family identification}

{We developed several complementary methods to identify asteroid families. The first one consists in
visualizing the distribution of proper elements in 3D. Given that no new family is seen to cross the
3:1 ($a=2.5$ au), 5:2 ($a=2.825$ au) and 7:3 ($a=2.958$ au) orbital resonances with Jupiter, we divide
the new catalog into four parts: the inner belt ($a<2.5$ au, including Hungarias), middle belt
($2.5<a<2.825$ au), pristine zone ($2.825<a<2.958$ au) and outer belt ($2.958<a<3.8$ au, excluding
Hildas). The search for new families is conducted separately in each of these zones. Two different
3D visualization codes were developed by us (D.N. and M.B.). The first code interactively displays the
distribution of proper orbits in a selected zone, allows the user to zoom out and zoom in, and
perform any kind of active rotation. The rotation is particularly useful because the user can easily check
whether any concentration seen in a projection is a real concentration of proper orbits in 3D.
The second code displays the $(e_{\rm p},\sin i_{\rm p})$ projection either in a narrow range of
$a_{\rm p}$ or with the $a_{\rm p}$ values coded in each dot's color. The dot's size is set inversely
proportional to the corresponding object's absolute magnitude which allows the user to identify
candidate families that stand out from the background of small and/or large objects.

For each candidate family we identify the lowest numbered asteroid that appears to be associated with
the family and use it as the family label. This association is not unique in many cases where there are
two or more large bodies in/near the family, with some being more/less offset from the family center.
Starting from the labeled bodies we then proceed by identifying each family with the Hierarchical Clustering
Method (HCM; Zappal\`a et al. 1990) and the usual metric
\begin{equation}
d = {3 \times 10^4\, {\rm m/s} \over \sqrt{a_{\rm p}}} \sqrt{ {5 \over 4} 
\left({\delta a_{\rm p}/a_{\rm p}}\right)^2 + 2 (\delta e_{\rm p})^2 + 2 (\delta \sin i_{\rm p})^2}\ , 
\end{equation}
where $3 \times 10^4\, {\rm m/s}$ is the orbital speed at 1 au and $\delta$ indicates differences
in the proper elements between two neighbor bodies. In each case we test many cutoff distances,
$d_{\rm cut}$, and visualize the results with the interactive software described above.\footnote{HCM
  clusters bodies by linking them together in a chain where the length of each segment is required to be
  $d < d_{\rm cut}$.}
This allows us to understand how the $d_{\rm cut}$ value needs to be adjusted to identify the whole family
seen in 3D. We pay particular attention to the extension of each candidate family in the proper semimajor
axis as the orbital resonances can create gaps in the distribution of proper elements and this
could artificially divide a family into two or more parts.

D.N. and M.B., working independently, produced two independent lists of candidate families. We then worked
together to resolve any differences. The two lists were largely overlapping but there were also $\sim$10\% of 
cases where one of us was more conservative in his approach. We accessed the likelihood that each
disputed concentration makes sense from what we know about the formation and dynamical evolution of asteroid
families (e.g., Nesvorn\'y et al. 2015). We also used the SDSS colors (Ivezi\'c et al. 2001, Parker et al. 2008)
and WISE albedos (Mainzer et al. 2011, Nugent et al. 2015)
to check whether identified candidate families are spectroscopically homogeneous,
and if so, whether they stand out from colors/albedos of the local background.

The identification method described above is subjective. It uses scientists' ability to identify concentrations
in the 3D distribution of proper elements, and their best judgment to establish whether a perceived
concentration does or does not constitute a real asteroid family (i.e. fragments produced by a disruptive collision
or spin-up driven fission). We believe that this approach is more powerful and reliable than any of the
mathematical and presumably more objective methods proposed elsewhere.\footnote{The advantage of mathematical
  algorithms is that their results are exactly reproducible whereas the ones involving subjective human
  intervention are not.}
For example, the method based on the
concept of the so-called Quasi Random Level (QRL; Zappal\`a et al., 1994) uses the asteroid population in the local
background (e.g., in the middle belt) to establish the likelihood that random statistical fluctuations would
produce any observed concentration. There are at least two problems with this method. 
First, the number density in proper element space is highly variable due to the primordial sculpting of the main belt
and resonances (e.g., Minton \& Malhotra, 2009). Applying the same QRL value in different parts of the main belt 
can therefore lead to unsatisfactory results (see Nesvorn\'y et al 2015, Section 7.3, for examples).
Second, asteroid families do not live in isolation but are frequently close
to each other, overlap, and/or are surrounded by empty regions. This introduces an ambiguity in the QRL definition,
because it is not clear a priori what region in $(a_{\rm p},e_{\rm P},i_{\rm P})$ space should be considered
to define the local QRL in the first place; results depend on this choice.

We used two methods to establish the statistical significance of newly identified candidate families. The first
method was proposed in Nesvorn\'y et al. (2002) to demonstrate the high statistical significance of the Karin family, which
is embedded in the much larger Koronis family. For that Nesvorn\'y et al. (2002) generated thousand mock orbital
distributions corresponding to the Koronis family and applied the HCM to each one. With $d_{\rm cut} = 10$ m/s,
no concentrations in this input were found containing more than a few dozen members, while the Karin family 
had 541 known members in 2015 (Nesvorn\'y et al. 2015). Therefore, the Karin family is significant at a (much)
greater than the 99.9\% level. The second method was borrowed from Rozehnal et al. (2016). We created three boxes
in the neighborhood of a family, one box centered at the family and the other two boxes below and above the family
in $\sin i_{\rm p}$; the boxes have the same shape and volume in $a_{\rm p},e_{\rm P},i_{\rm P}$. We counted
how many asteroids there were in each box, defining $N_1$, $N_2$ and $N_3$, where $N_2$ is the number in
the middle box containing the candidate family, generated $n=10^7$ random distributions with $N=N_1+N_2+N_3$
bodies in the region covered by all three boxes, and counted the number of positive trials, $n^+$, for which the
number of randomly generated bodies in the middle box equaled or exceeded $N_2$.
The probability that the observed number of bodies in the middle box is a result of random fluctuations is then
$n^+/n$, and the statistical significance of the observed family is $1-n^+/n$. Only families with high statistical
significances were included in the final catalog.} 

\section{Catalog}

\begin{figure}
\epsscale{1.2}
%\hspace*{-1.7cm}\plotone{proper1.pdf}
\hspace*{-1.7cm}\plotone{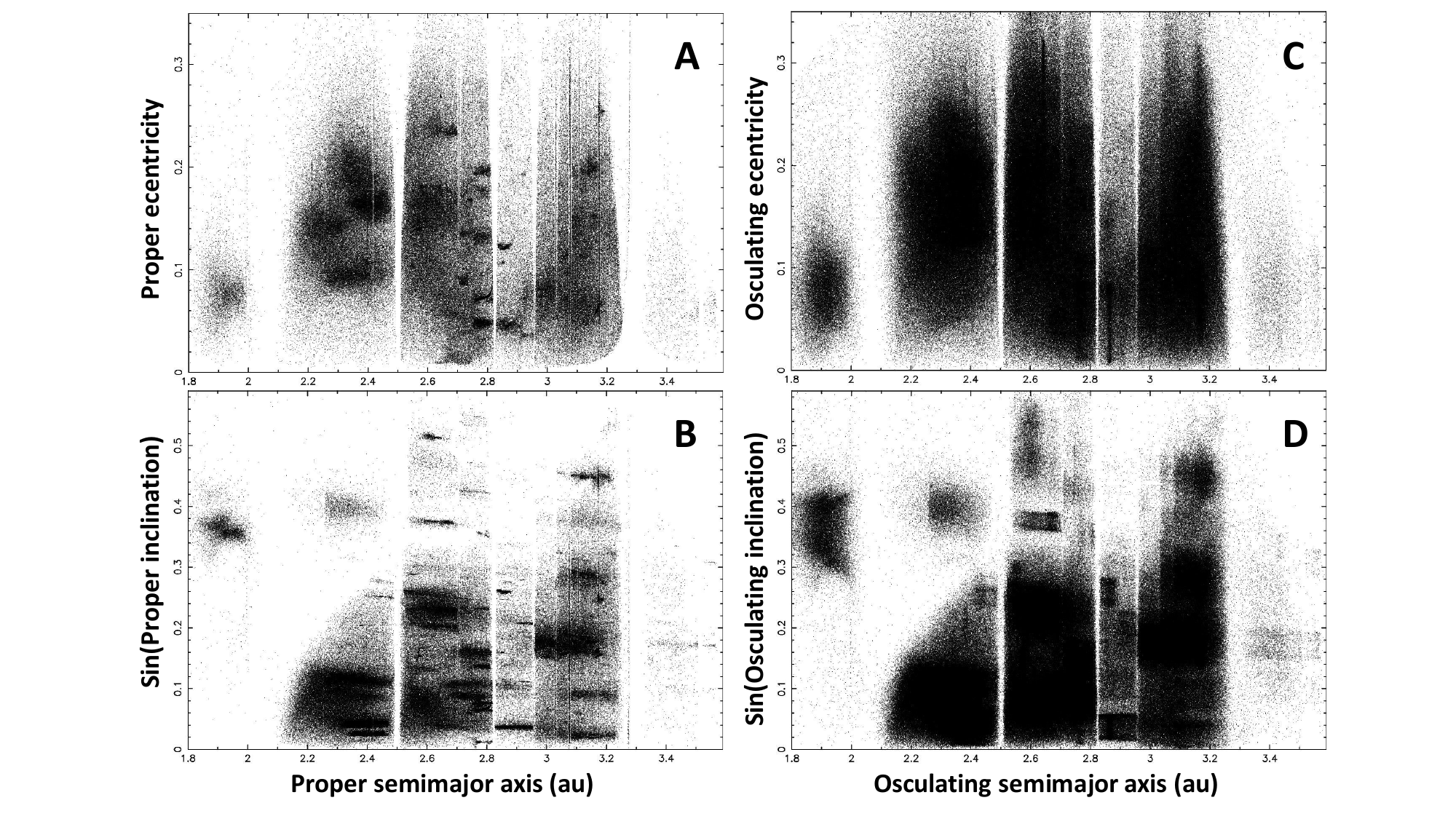}
\caption{Proper (panels A and B) and osculating orbits (panels C and D) of main belt asteroids. 
In the proper element space, various orbital structures come into focus. }
\label{proper1}
\end{figure}

The new catalog of proper elements is available at \texttt{https://asteroids.on.br/appeal/},
\texttt{www.boulder.swri.edu/\~{}davidn/Proper24/}, and the PDS
node.\footnote{\texttt{https://pds.nasa.gov/}}  
Here we give a brief description of the catalog content. There are 1,249,051 main belt asteroids in total
(Fig. \ref{proper1}). Figures \ref{proper1}A-B show the proper orbits from the new catalog.
It is clear from these figures that the asteroid belt shows intricate orbital structure (asteroid 
families, resonances, etc.), which is not obvious from the distribution of osculating elements 
(Figs. \ref{proper1}C-D). That is, in fact, the chief motivation behind calculating the 
proper elements: in the proper element space, various orbital structures come into focus (e.g.,
Milani \& Kne\v{z}evi\'c 1994).    
We do not give proper elements for unstable bodies that were eliminated from the integration in 10 Myr 
(because they impacted one of the planets, the Sun, or were ejected from the solar system).
The ASCII file \texttt{eliminated\_bodies.dat} lists all eliminated bodies, the time of 
elimination, and the elimination flag (0 -- impact on the Sun, 1 to 8 -- planet impacts, 1 is 
Mercury, 8 is Neptune, 9 - ejection from the solar system). 

The proper element catalog (\texttt{proper\_catalog24.dat}) is an ASCII file with 
11 columns. Table 2 shows the first 20 lines of the catalog. The three proper elements are 
given in columns 1 ($a_{\rm p}$), 3 ($e_{\rm p}$) and 5 ($\sin i_{\rm p}$) (columns 2, 4 and 6 list their 
uncertainties). The proper frequencies $g$ and $s$ are listed in columns 7 and 8 (uncertainties not 
given for brevity). The absolute magnitudes from MPC are in column 10. Column 11 can be used to select 
well-determined orbits with any number of oppositions.\footnote{The single-opposition orbits must be
  used with care. While in most cases the single-opposition orbits are good enough to obtain reliable
  proper elements, there are also instances where this is not the case. We leave the decision to the
  user who can easily apply any cut based on the values listed in column 11.}
The values given here were imported from the 
MPC catalog on February 9, 2024 (the source MPC catalog is included in the distribution). The proper 
element package available on the PDS node also contains the modified \texttt{swift\_rmvs4} code that 
was used to integrate orbits and all tools that were used for analysis.\footnote{The FMFT code is also 
available at \texttt{https://www.boulder.swri.edu/\~{}davidn/fmft/fmft.html}}

\begin{figure}
\epsscale{0.8}
%\plotone{errors.eps}
\plotone{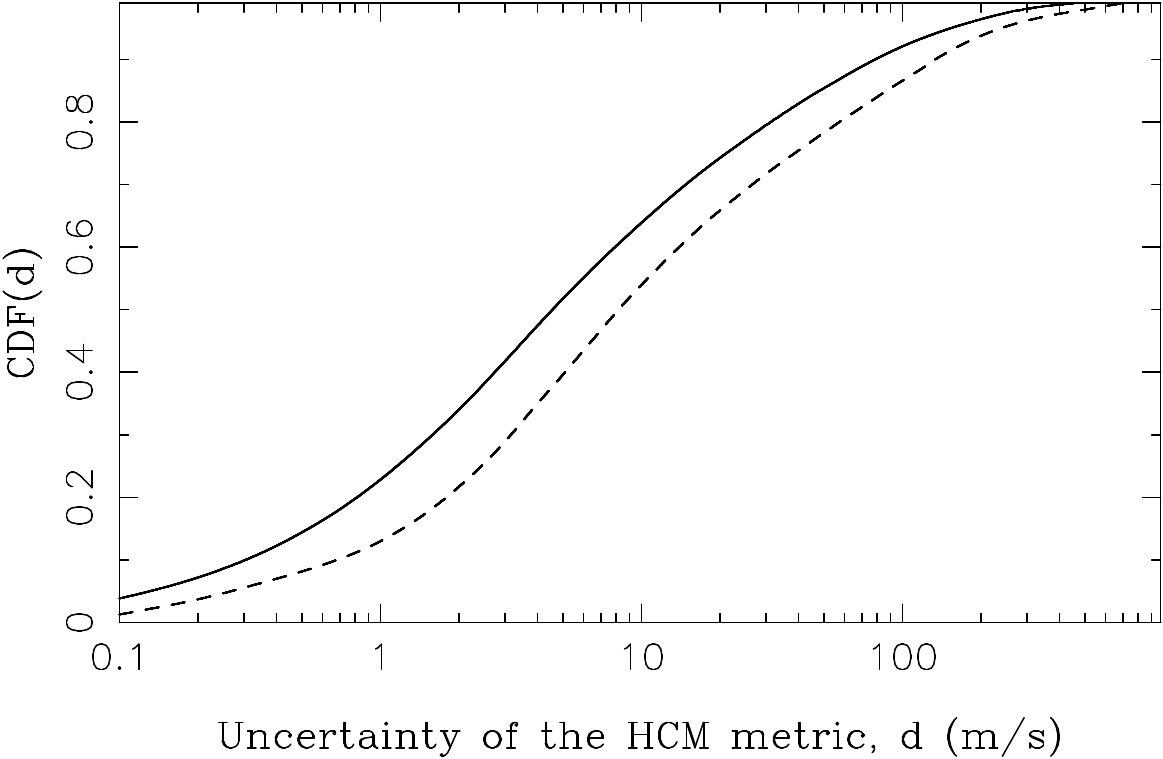}
\caption{Uncertainties of proper elements in terms of the $d$ metric: the new catalog (solid line), Novakovi\'c's AFP 
catalog (dashed line).}
\label{error}
\end{figure}

\begin{figure}
\epsscale{1.8}
%\hspace*{0.0cm}\plotone{proper2.pdf}
\hspace*{0.0cm}\plotone{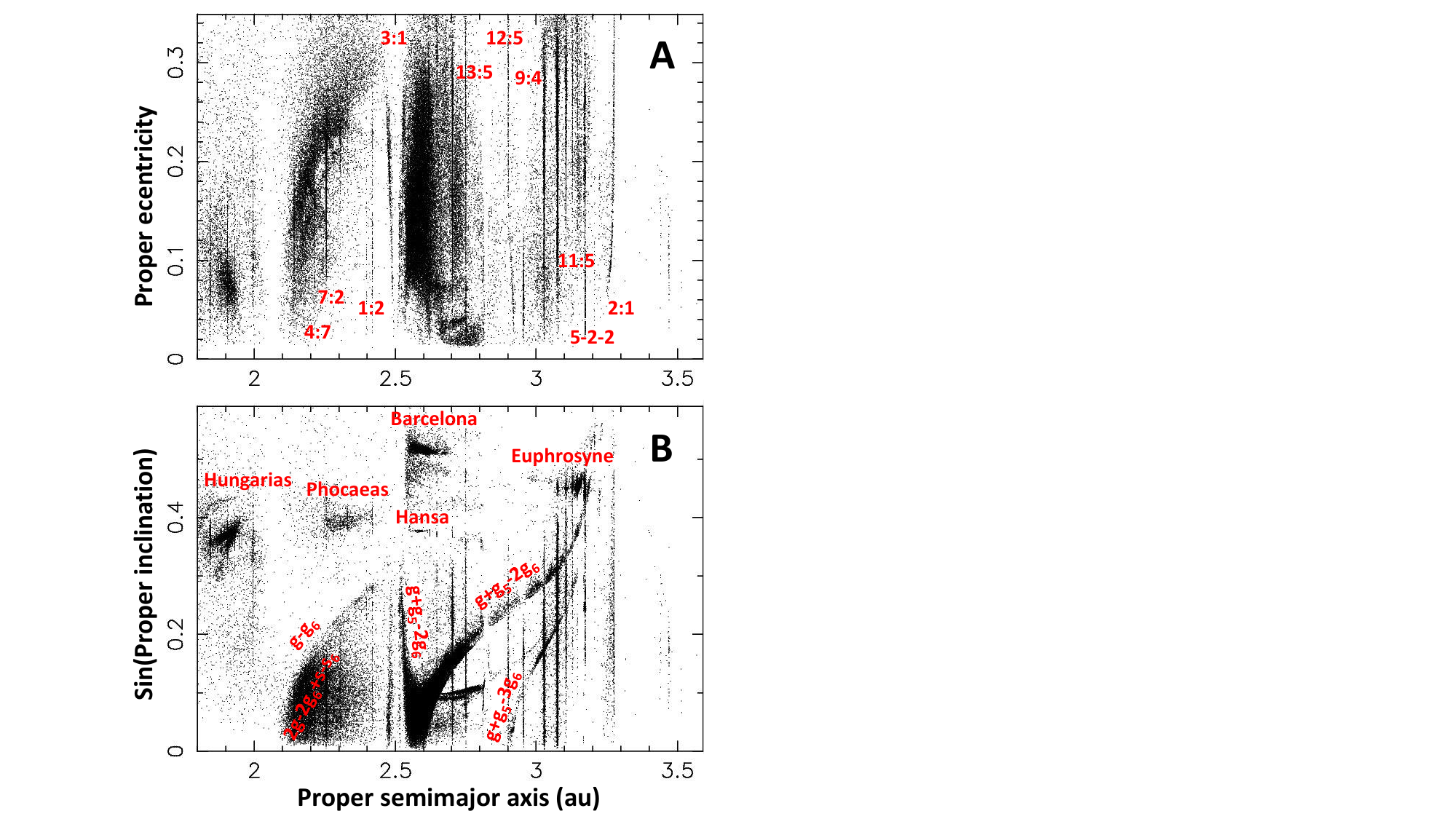}
\caption{The proper orbits of main belt asteroids with relatively large errors ($d>100$ m/s). Problematic 
mean motion resonances with Jupiter and Mars, secular resonances, and asteroid groups/families are labeled.}
\label{error2}
\end{figure}

The uncertainties of three proper elements, $\delta a_{\rm p}$, $\delta e_{\rm p}$ and
$\delta \sin i_{\rm p}$, can be converted to a single number, $d$, defined by Eq. (1),
where $d$ is the usual metric used for asteroid family identification with HCM (Zappal\`a et al. 1990). 
The cumulative distribution of $d$ is shown in Fig. \ref{error}. We achieve a slightly better 
precision than the AFP catalog (Novakovi\'c et al. 2022) but the difference is not large. In about 64\% of cases the precision 
is better than 10 m/s, which should be satisfactory for the identification of even very compact 
families.\footnote{The HCM cutoff used in the identification of the most compact asteroid families
  is $d_{\rm cut}\simeq 10$ m/s. Thus, even the most compact asteroid families can be identified
  if the precision of proper elements is better than 10 m/s.}
In about 7\% of cases, the error exceeds 100 m/s. Figure \ref{error2} shows that this 
typically happens close to specific mean motion and secular resonances. There are several 
families with high inclinations (e.g., Hansa, Barcelona) that are affected by these large errors. 

\begin{figure}
\epsscale{1.0}
%\plotone{veritas1.eps}\\[1.cm]
%\plotone{veritas2.eps}
\plotone{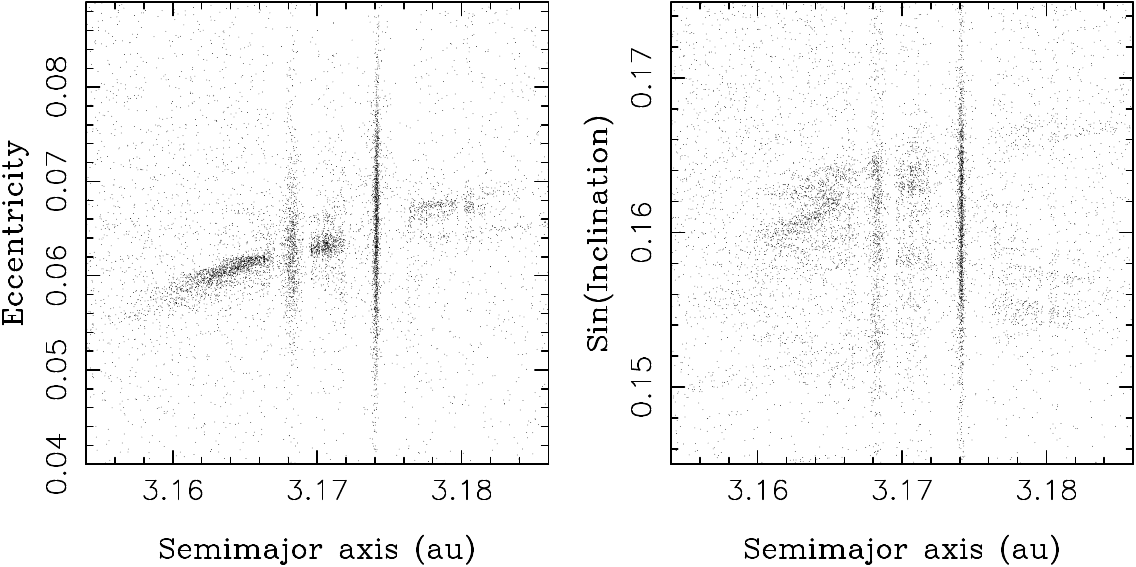}\\[1.cm]
\plotone{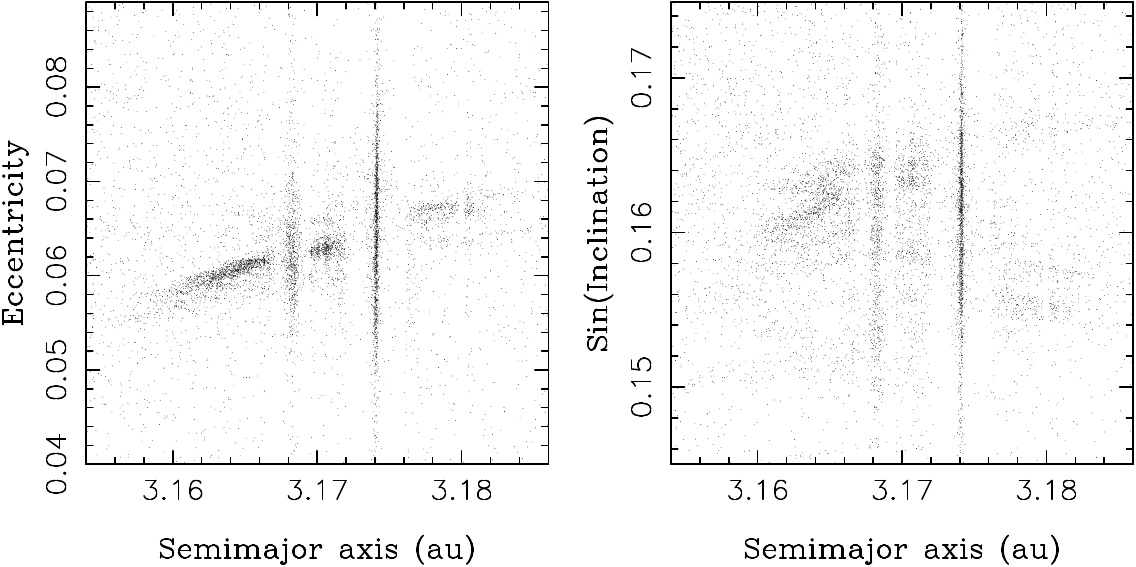}
\caption{The dynamical structure of the Veritas family: the proper elements computed here
(top panels) and the proper elements from \texttt{http://asteroids.matf.bg.ac.rs/fam/} 
(bottom panels).}
\label{veritas}
\end{figure}

In the great majority of cases, however, the precision of the new catalog is good enough to
identify known families and recover their intricate dynamical structure. Here we illustrate this
on the Veritas family, a well-known 
family in the outer belt that has been identified as the source of late-Miocene dust shower 
on the Earth (Farley et al. 2006). Figure \ref{veritas} compares our proper elements with
the AFP catalog (Novakovi\'c et al. 2022). The two distributions are very similar. Our 
catalog has roughly 20\% more asteroids than the AFP catalog,  which allows some of the 
subtle features to stand out slightly more clearly. For example, in the left panels of
Fig. \ref{veritas}, there are two streaks for $a_{\rm p}>3.176$ au, one with slightly higher 
and one with slightly lower proper eccentricities, that appear to diverge from each other
with increasing semimajor axis (these streaks were already visible in the AFP data).
The upper streak with $e_{\rm p}\simeq0.065$--0.07 has 
lower proper inclinations ($\sin i_{\rm p}\simeq0.154$--0.158). This may tell us something
interesting about the velocity field of fragments from the site of the original breakup
(e.g., Carruba et al. 2016). The
vertical features near 3.168 au and 3.174 au are mean motion resonances ((3 3 -2) and 
(5 -2 -2), respectively; Tsiganis et al. 2007).       

\section{New asteroid families}

\begin{figure}
\epsscale{1.0}
%\plotone{fam1.eps}
\plotone{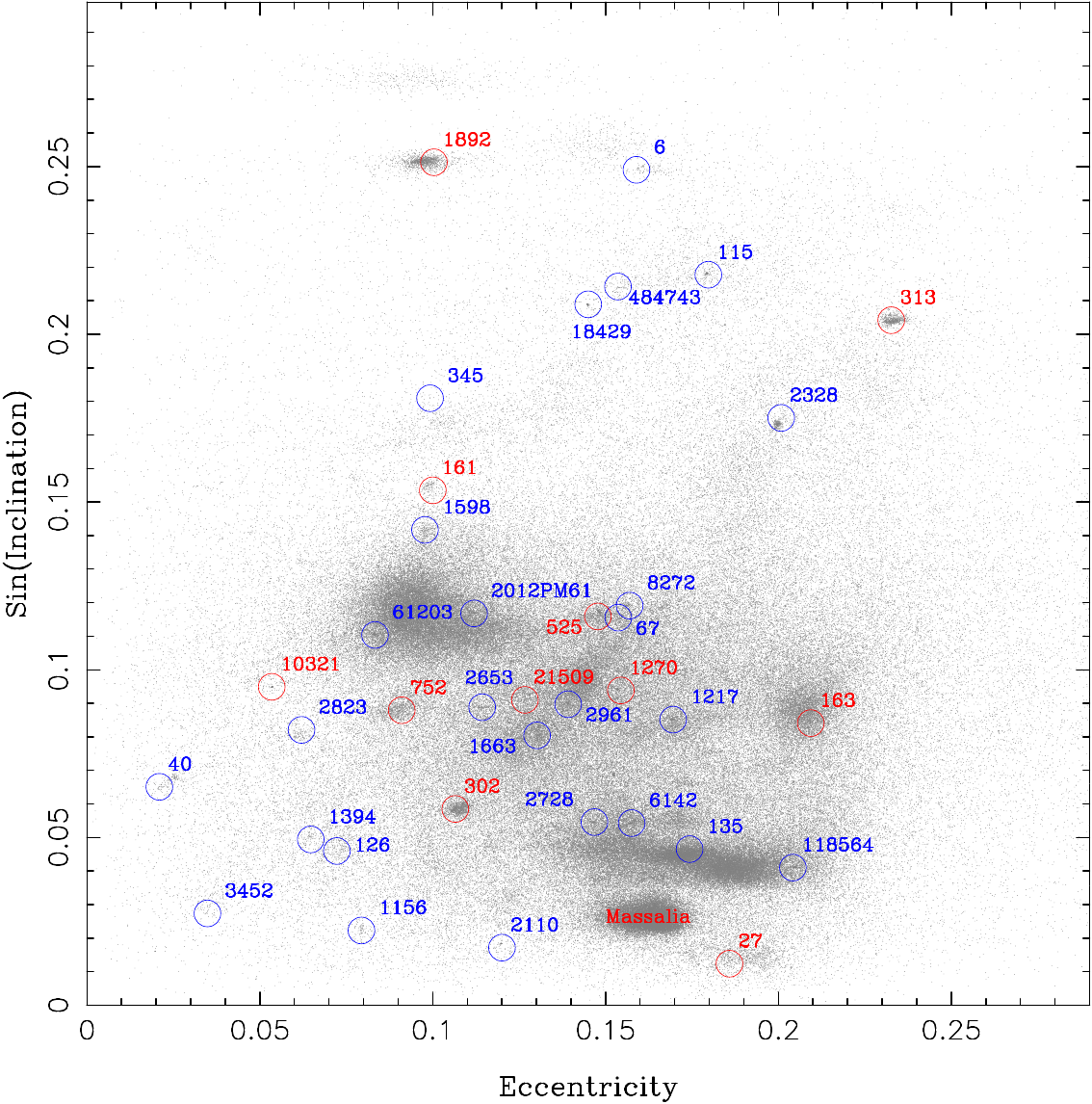}
\caption{Proper eccentricities and sine of proper inclinations for inner belt asteroids 
($a_{\rm p}<2.5$ au). Notable families are labeled: blue labels stand for families identified 
  in this work and not listed in Nesvorn\'y et al. (2015), red labels are families known
  previously (not all previously known families are labeled).}
\label{fam1}
\end{figure}

\begin{figure}
\epsscale{1.0}
%\plotone{fam2.eps}
\plotone{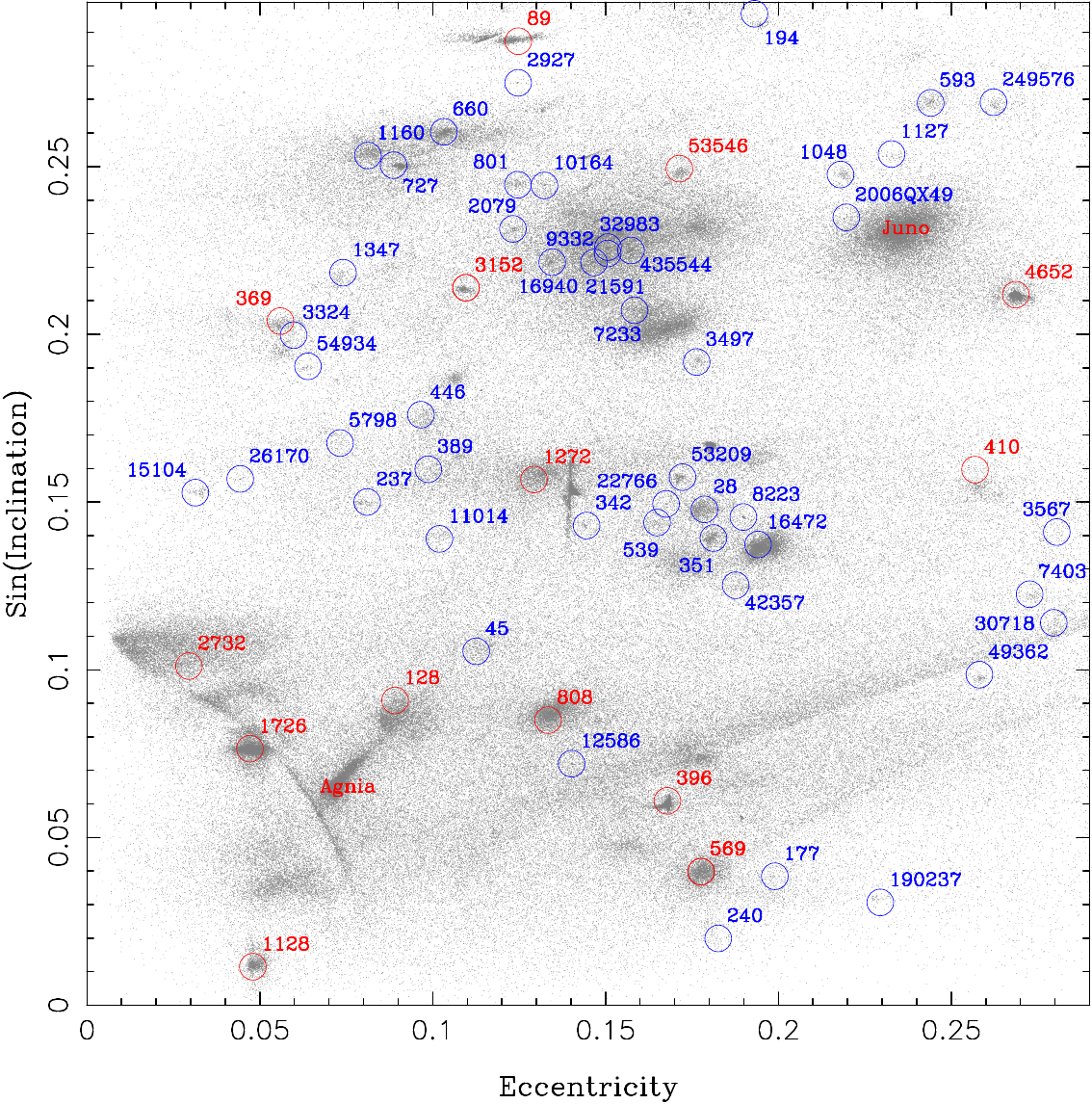}
\caption{Proper eccentricities and sine of proper inclinations for middle belt asteroids 
($2.5<a_{\rm p}<2.825$ au). Notable families are labeled: blue labels stand for families identified 
  in this work and not listed in Nesvorn\'y et al. (2015), red labels are families known previously
  (not all previously known families are labeled). }
\label{fam2}
\end{figure}

\begin{figure}
\epsscale{1.0}
%\plotone{fam3.eps}
\plotone{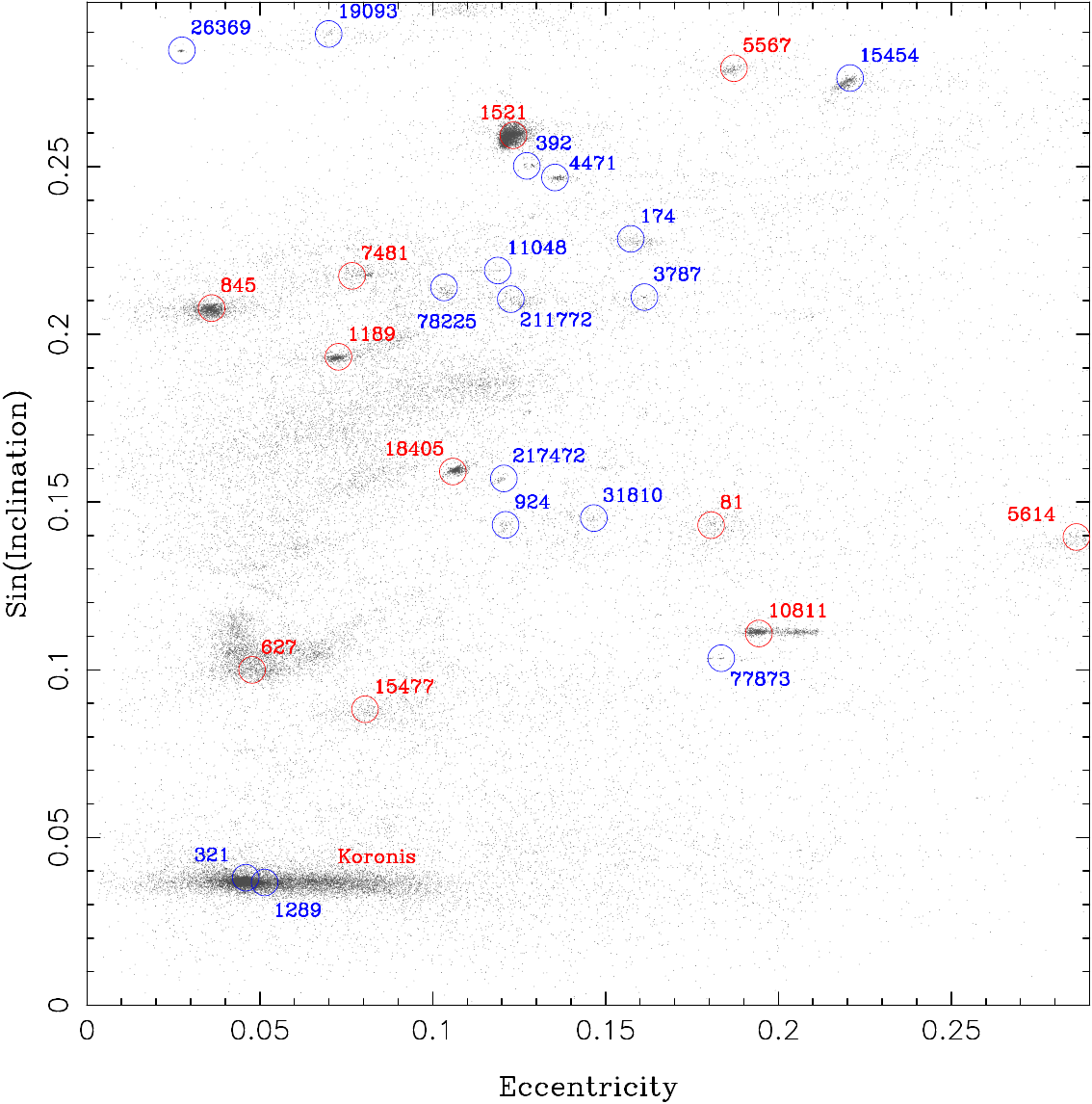}
\caption{Proper eccentricities and sine of proper inclinations for asteroids in the pristine zone 
($2.825<a_{\rm p}<2.958$ au). Notable families are labeled: blue labels stand for families identified 
  in this work and not listed in Nesvorn\'y et al. (2015), red labels are families known previously
  (not all previously known families are labeled).}
\label{fam3}
\end{figure}

\begin{figure}
\epsscale{1.0}
%\plotone{fam4.eps}
\plotone{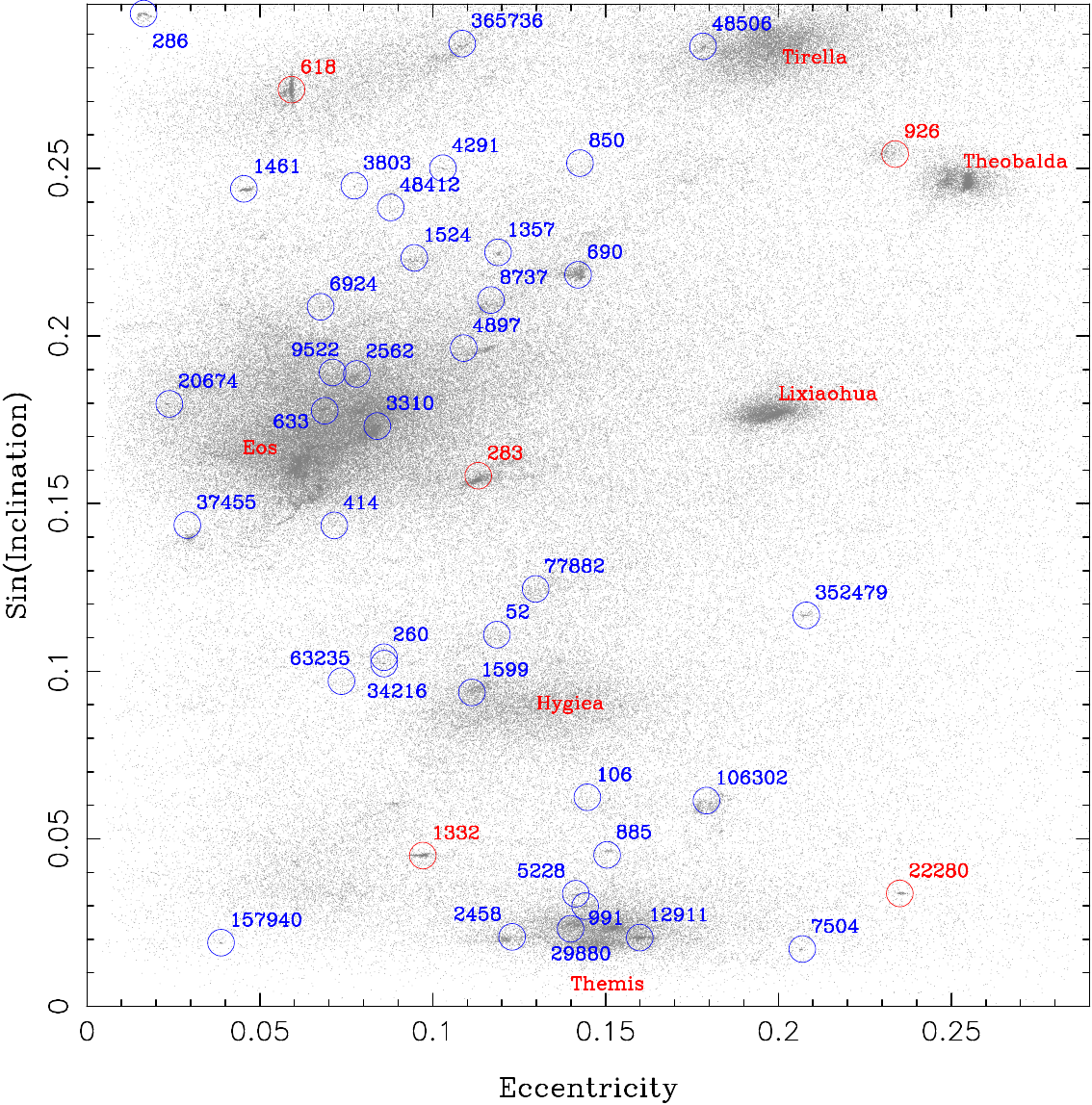}
\caption{Proper eccentricities and sine of proper inclinations for outer belt asteroids
($2.958<a_{\rm p}<3.8$ au). Notable families are labeled: blue labels stand for families identified 
  in this work and not listed in Nesvorn\'y et al. (2015), red labels are families known previously
  (not all previously known families are labeled). }
\label{fam4}
\end{figure}

Figures \ref{fam1}--\ref{fam4} show the orbital distribution of asteroids in the inner
($a<2.5$ au), middle ($2.5<a<2.825$ au), pristine ($2.825<a<2.958$ au) and outer parts
($a>2.958$ au) of the main belt. We used the methods described in Sect. 2.2 to conduct
a search for new asteroid families. All together, we found 153 statistically significant
families (see below) that have not been reported in Nesvorn\'y et al. (2015) (Tables 3-7).
A cross-check against recent publications revealed that 17 of these families were
found previously (Novakovi\'c et al. 2011, 2014, 2022; Carruba et al. 2015, 2019;
Dykhuis et al. 2015; Novakovi\'c \& Radovi\'c 2019; Tsirvoulis 2019; Bro\v{z} et al. 2024)
and/or are listed in the AFP catalog (Novakovi\'c et al. 2022). There are 136 new families:
28 in the inner belt (including Phocaeas)\footnote{There are five new families in Phocaeas:
Bezovec, Chesneau, Bascom, 1999 XB232 and 1999 RX33.}, 47 in the middle belt,
15 in the pristine zone, 33 in the outer belt (including Cybeles but excluding
Hildas)\footnote{There is 8 families in Cybeles that were not reported in Nesvorn\'y et al.
  (2015), 6 with low inclinations ($\sin i_{\rm p} < 0.3$), Huberta, Liriope, Fukui, Schlichting,
  2001 BV20 and 2010 WK8, and 2 with high inclinations ($\sin i_{\rm p} > 0.3$), Helga and
  Abastumani. Five of these Cybele families were already identified in Carruba et al.
  (2015, 2019).}, and 13 with high orbital inclinations ($\sin i_{\rm p} > 0.3$; Table 7).
We used the SDSS colors (Ivezi\'c et al. 2001, Parker et al. 2008) and WISE albedos (Mainzer
et al. 2011, Nugent et al. 2015) to determine the taxonomic type of many new families (Tables 3-7).

There were 122 asteroid families reported in Nesvorn\'y et al. (2015). The 153 families identified here
therefore represent a 126\% increase over the catalog published in Nesvorn\'y et al. (2015).
One of the families reported in  Nesvorn\'y et al. (2015), (709) Fringilla (FIN 623), was split
into two overlapping families, (19093) 1979MM3 and (37981) 1998HD130. There are now 274 known
families in the asteroid belt in total (families in the resonant Hilda population are not
counted here). 

{In a great majority of cases, the new families clearly stand out from the background such that their
significance is undisputed. We first illustrate the methods described in Sect. 2.2 for
the (3787) Aivazovskij  (pristine zone) and (4291) Kodaihasu (outer belt) families. These new
families are compact but not exceedingly so; they represent a typical
case of compact families identified here. (3787) Aivazovskij has 12 members identified with
$d_{\rm cut}=15$ m/s, (4291) Kodaihasu has 16 members identified with $d_{\rm cut}=20$ m/s. 
Applying the HCM-based method described in Sect. 2.2, we establish that both families are
significant at least at the 99.9\% level. In the case of (3787) Aivazovskij, there are 12 bodies
in the box around the family ($N_2=12$), and no bodies in boxes directly below or above ($N=12$).
The probability that this happens by chance is only $\simeq 2.6\times10^{-6}$. In the case of (4291) Kodaihasu,
there are 16 bodies in the box around the family ($N_2=16$), and no bodies in boxes directly
below or above ($N=16$). The probability that this happens by chance is only $\simeq 3\times10^{-8}$
($10^8$ trials used here). This demonstrates that the two families are statistically significant. 

We applied the box method (Sect. 2.2, Rozehnal et al. 2016) to all families identified here. In 137 cases,
representing $\simeq 90$\% of the total of 153, the number of trials ($10^7$) was insufficient to distinguish
the statistical significance from 1. This shows that all these families are significant at least at
the 5-sigma level. Table 8 shows the statistical significance for the remaining 16 families.
Most of these families are compact and contain a small number of members; these small families are apparently
more likely to be produced by statistical fluctuations. Still, in most cases, the probabilities reported
in Table 8 are comfortably small. In addition, many of the compact families show clustered
orbital longitudes and past convergence (see below); this includes the family around (9332) 1990SB1
(Table 8).
The statistical significance of these families would greatly increase if these properties were taken
into account.

The three most problematic families are: (240) Vanadis (17 members, 0.003 probability), (22766) 1999AE7
(5 members, 0.01 probability) and (77882) 2001 SV124 (316 members, 0.05 probability). We prefer
to report these families here but acknowledge that these cases need a more detailed analysis and/or
confirmation from additional data. It is useful to include them here to see whether these borderline
cases will or will not be confirmed with new data. The (77882) 2001 SV124 family is one of only two
larger families (with more than 20 members) -- the other one being (106) Dione -- that are listed
in Table 8.}  

We now comment on several new families of special interest. 
Notably, the family around (1461) Jean-Jacques in the outer belt is interesting because 
this object, 30-40 km in diameter, is classified as a metallic M-type asteroid in Tholen 
taxonomy. Another interesting case is a new family in the inner belt near asteroids (2110)
Moore-Sitterly and (44612) 1999 RP27. This family has 17 members that show obvious
clustering in the perihelion and nodal longitudes (Table 9). Pravec \& Vokrouhlick\'y
(2009) (see also Pravec et al. 2010), identified the two largest bodies in the family,
2110 and 44612, as an asteroid pair (Vokrouhlick\'y \& Nesvorn\'y 2008). Polishook et al.
(2014a,b) found their spectral classification as intermediate between S and Sq. They have rotation 
periods 3.345 hr and 4.907 hr, respectively, and both are retrograde rotators with poles
pointing to ecliptic latitudes between $-70^\circ$ and $-80^\circ$ (Pravec et al. 2010, 
2019; Polishook 2014).

\begin{figure}
\epsscale{0.8}
%\plotone{moore1.eps}
\plotone{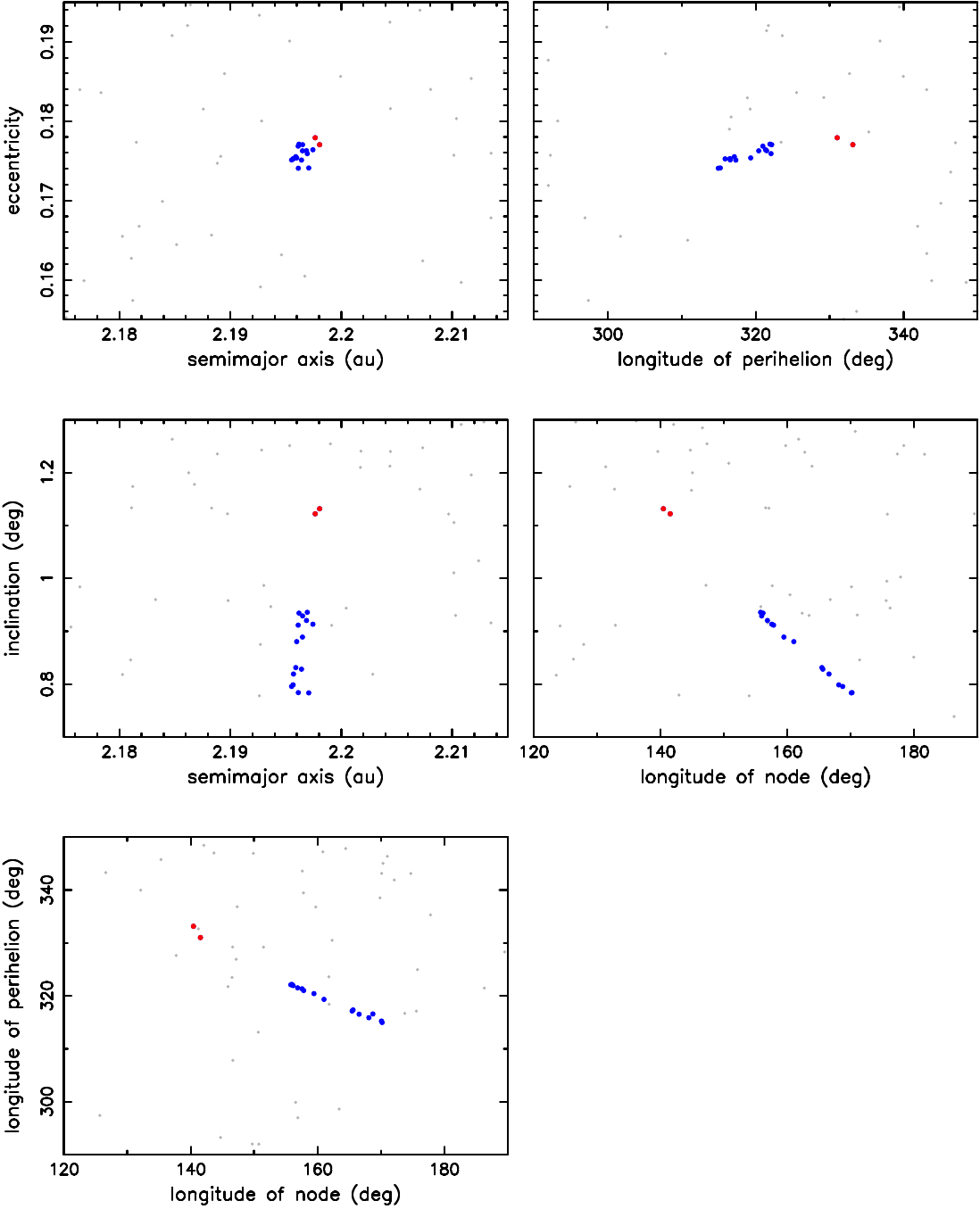}
\caption{The osculating orbits of asteroids near the Moore-Sitterly family (MJD epoch 60400.0):
(2110) Moore-Sitterly and (44612) 1999 RP27 are shown by red dots, other family members by 
blue dots; grey dots show background asteroids.}
\label{moore1}
\end{figure}

\begin{figure}
\epsscale{0.8}
\plotone{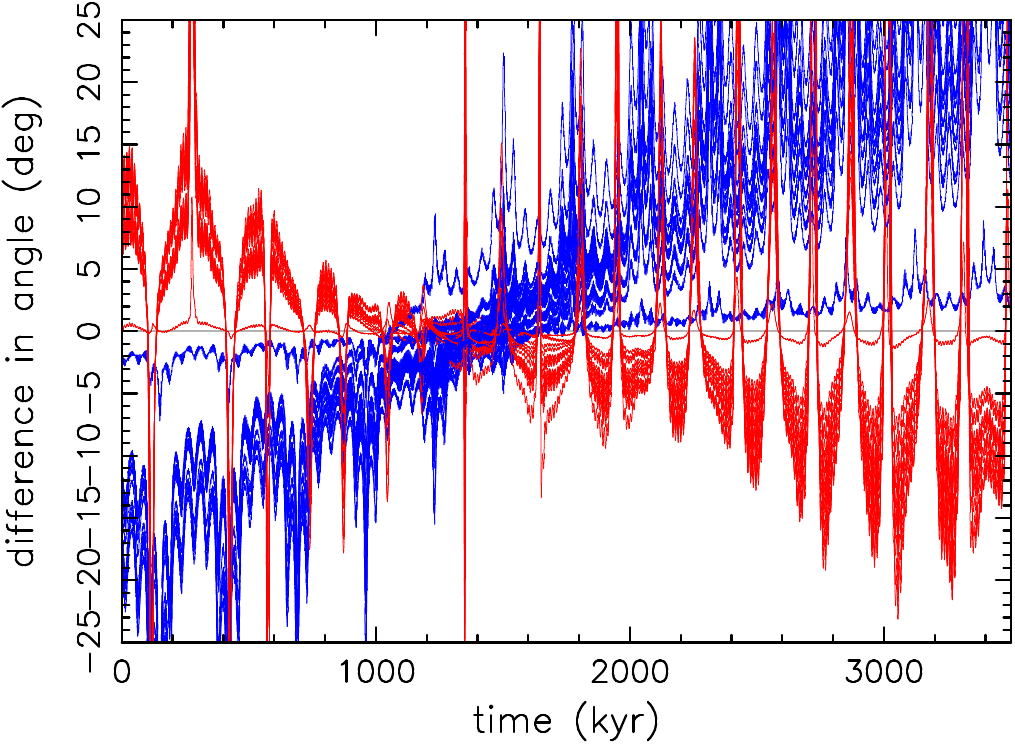}
\caption{The past orbital convergence of secular angles 1.2-1.5 Myr ago indicates that 
the Moore-Sitterly family must be very young. The nodal longitudes are shown by red line,
perihelion longitudes by blue. All 17 known family members are plotted here. 
Note that this result was obtained without considering
the Yarkovsky acceleration in the $N$-body integrator. A careful analysis with the 
Yarkovsky force is left for future work.}
\label{moore2}
\end{figure}

Figure \ref{moore1} shows osculating orbits of the Moore-Sitterly family members. The two largest bodies,
2110 and 44612, have slightly larger semimajor axis values than the rest of the family. There
is an obvious orbital convergence of secular angles 1.2-1.5 Myr ago (Fig. \ref{moore2}), which 
clearly demonstrates that this family must be very young. We find that the semimajor axis difference
between 2110 and 44612 corresponds to the ejection velocity of $\sim$ 2 m/s, which is
comparable to the escape velocity from (2110) Moore-Sitterly. Some of the small fragments
are $\sim$ 10 times farther away, which would indicate an ejection velocity of $\sim$ 20 m/s -- 
much larger than the escape velocity -- and this would be puzzling. If all small fragments were
retrograde rotators, however, their displacement could be explained by the Yarkovsky
drift sunward from their original locations over 1.2-1.5 Myr. The magnitude distribution of the 
family members (Table 9) indicates a notable separation between the two largest members (formerly 
found to constitute a pair of asteroids), and the smaller fragments. This is reminiscent of the
Hobson family case (e.g., Vokrouhlick\'y et al. 2021), where the authors speculated that the parent 
body of the Hobson family was a binary asteroid. Indeed, the nearly equal-size binaries should 
represent some 15\% of $D<10$ km main belt asteroids (e.g., Pravec et al. 2016), and it is inevitable 
that some of the binaries give birth to families.

These same analysis was applied all families listed as ''compact'' in Tables 3-7. We do not comment
on these cases in detail here, this will be the subject of a separate publication, but point out that 
the new families (1217) Maximiliana, (6084) Bascom, (10164) Akusekijima and (70208) 1999RX33 show
past convergence of orbital longitudes, and are estimated here to have formed 0.5-2.5 Myr ago. 

\begin{figure}
\epsscale{0.8}
%\plotone{9332_ele.eps}
\plotone{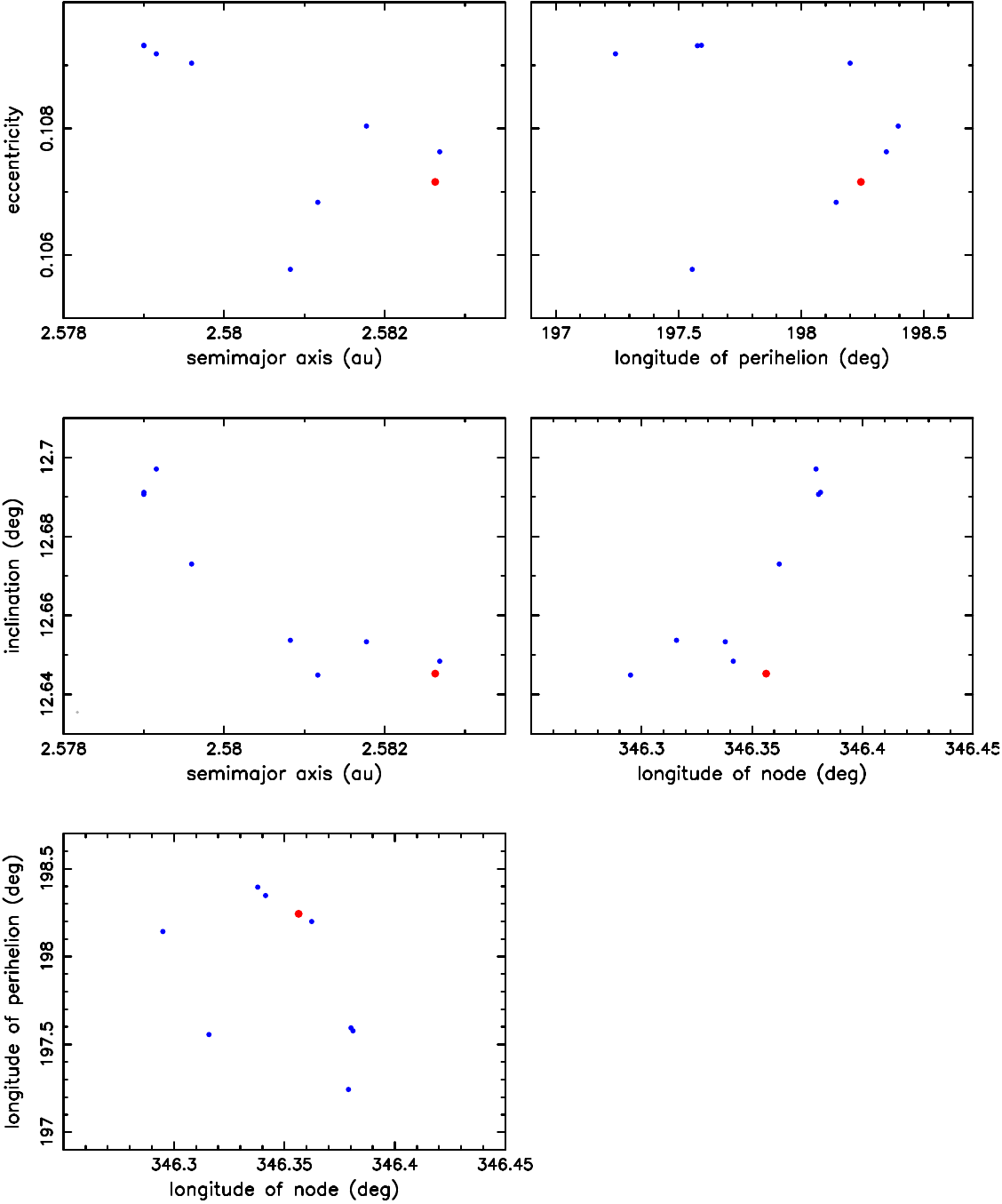}
\caption{The osculating orbits of asteroids near the (9332) 1990SB1 family (MJD epoch 60400.0):
the red dot is 9332, blue dots are small family fragments. The family is so compact that no 
background asteroids appear in any of these projections. Two family members, 2016DB45 and 
2016EQ6, have practically identical orbits; the difference in the mean anomaly is only 
0.6 deg. They appear as two overlapping dots in all projections.}
\label{9332a}
\end{figure}

\begin{figure}
\epsscale{0.8}
%\plotone{9332_zoom.eps}
\plotone{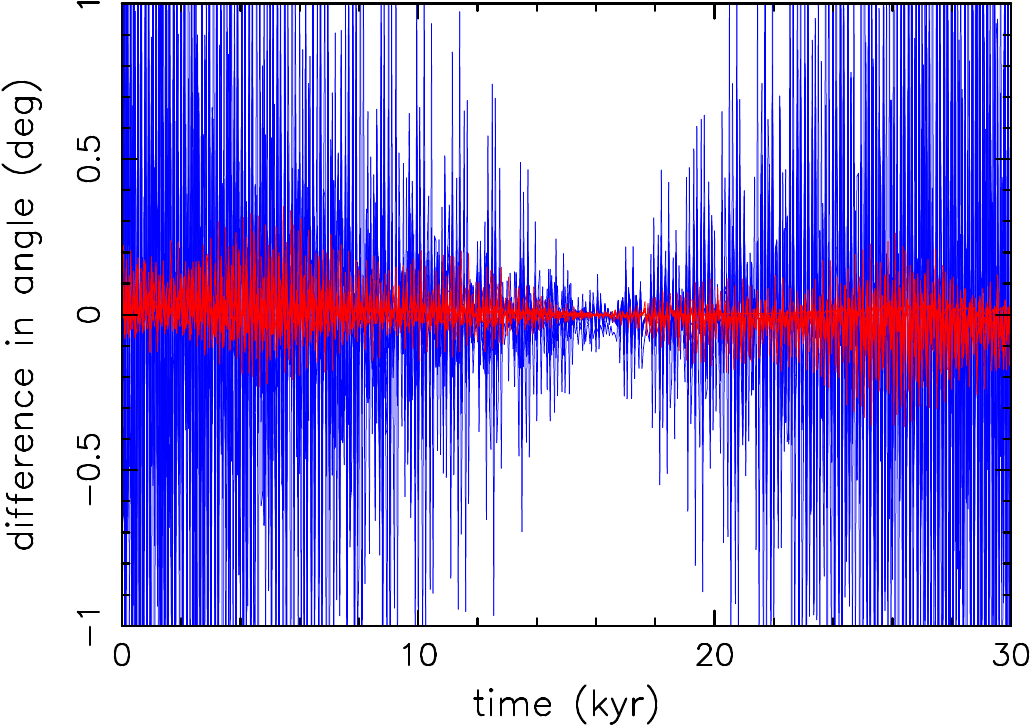}
\caption{The past orbital convergence of secular angles 16-17 kyr ago indicates that 
the (9332) 1990SB1 family is extremely young. The nodal longitudes are shown by red 
lines, perihelion longitudes by blue. The mean anomalies of family members, not shown here,
converge at 16-17 kyr ago as well. All 9 known family members are plotted here. 
This result was obtained without considering the Yarkovsky acceleration in the $N$-body integrator.
A careful analysis with the Yarkovsky force is left for future work.}
\label{9332b}
\end{figure}

We also identify the youngest asteroid family found in the main belt so far. Asteroid (9332) 
1990SB1 with $H = 13.2$ ($D=6.8$ km for a reference albedo $p_{\rm V}=0.2$), $a = 2.583$ au, 
$e = 0.107$, $i = 12.65$ deg (osculating elements), is a member of the large Eunomia family
in the middle main belt (Nesvorn\'y et al. 2015). It has a relatively short rotation period, 
$2.987$ hr, and the lightcurve amplitude of 0.39 mag. Pravec et al (2019) found it is a binary; 
the satellite has an orbital period of 48.84 hr. Now, there are 8 small asteroids near (9332) 
1990SB1 with very similar orbits (Table 10 and Fig. \ref{9332a}). We integrated their orbits 
backward in time and established that the family formed only 16-17 kyr ago (Fig. \ref{9332b}). 
This is younger that the previous record holder, the (5026) Martes family (Vokrouhlick\'y
et al. 2024), with the estimated age of $\simeq 18.1$ kyr. Assuming that the (9332) 
1990SB1 family and satellite around (9332) 1990SB1 formed at the same time, we now know 
the age of the binary as well. This may be useful to understanding the binary formation, and 
evolution of binary orbits by tides and non-gravitational effects. We will report a more 
thorough analysis of this case in forthcoming publications.

\section{Conclusions}

The main findings of this work are summarized as follows:
\begin{enumerate}
\item The proper elements of asteroids are computed from the Fourier method (\v{S}idlichovsk\'y \&
  Nesvorn\'y 1996, Kne\v{z}evi\'c \& Milani 1999). We speed up and improve the computation by
  using the symplectic integrator/corrector and Frequency Modified Fourier Transform
  (\v{S}idlichovsk\'y \& Nesvorn\'y 1996).
\item The new catalog of proper elements for $1.25\times10^6$ main belt asteroids is available 
  at \texttt{https://asteroids.on.br/appeal/}, \texttt{www.boulder.swri.edu/\~{}davidn/Proper24/}, 
  and the Planetary Data System (PDS) node, \texttt{https://pds.nasa.gov/}.
\item A search for collisional families yielded 153 cases not reported in Nesvorn\'y et al. (2015):
  31 in the inner belt, 48 in the middle belt, 17 in the pristine zone, 39 in the outer belt, and
  18 in high inclinations. 136 of these cases are new discoveries and 17 were identified in various
  other publications. There are now 274 known families in the asteroid belt in total
  (the Hilda families are not included here). 
\item The (2110) Moore-Sitterly family is a close pair of relatively large bodies, 2110 and 44612,
  and 15 small members all located sunwards from 2110 and 44612, presumably a consequence of the Yarkovsky
  drift over the estimated family age (1.2-1.5 Myr).
\item New families (1217) Maximiliana, (6084) Bascom, (10164) Akusekijima and (70208) 1999RX33 are
  estimated to have formed 0.5-2.5 Myr ago. The new family around outer belt asteroid (1461) Jean-Jacques
  is interesting because this object is classified as a metallic M-type in the Tholen taxonomy.
\item The new family around middle belt asteroid (9332) 1990SB1 has 9 members. It is the youngest
  family found in the asteroid belt so far (estimated formation only 16-17 kyr ago). Asteroid (9332) 
  1990SB1 itself is a binary (Pravec et al 2019). 
\end{enumerate}
A systematic characterization of new asteroid families is left for future work.  

The methodology described here offers a practical method to compute asteroid proper elements
for the large amount of data expected from the Rubin Observatory (Schwamb et al. 2023). There are two
improvements. First, the symplectic integrator/corrector can speed up the computation of proper elements
by a factor of $\sim$5.\footnote{The beauty of symplectic integrators is that they can be used with
  time steps up to $\sim$ 1/20 of the shortest orbital period and that without any excessive growth of the
  integration error. Here we used a conservatively short time step, about 1/80 of the orbital period of
  Mercury, and verified that that the method works well -- with only a modest effect on the accuracy 
  of proper elements -- for time steps as long as 5 days. Compared to that, the multistep predictor in the
  ORBIT9 integrator uses a time step that is typically $\sim$ 1/100 of the shortest orbital period (Milani
  \& Nobili 1988); this would correspond to $\sim$ 21 hours for the orbital integration with Mercury. Therefore,
  since the ORBIT9 integrator needs to evaluate the gravitational accelerations up to $\sim 5$ times more
  often than the \texttt{Swift} integrator, the speed up can be as large as the factor of $\sim$ 5.}
The exact speed-up factor depends on the timestep used in the $N$-body integrator. Here we use a conservatively
short timestep (1.1 day) but point out that symplectic timesteps up to $\simeq 5$ days can yield
accurate proper elements. Second, the massive parallelization on 10,000+ cores of the NASA Pleiades
Supercomputer (as demonstrated here, {implemented via a PBS script; also see Novakovi\'c et al. 2009})
will allow us to compute the proper elements for $10^7$ main-belt asteroids in under a month of wall-clock time.

\section{Appendix A: Frequency splitting}

Here we illustrate the frequency splitting in the case of asteroid (5) Astraea. This issue has been
reported in \v{S}idlichovsk\'y \& Nesvorn\'y (1996) who found that (5) Astraea has a chaotic orbit near
the secular resonance $g + g_5 - 2 g_6 \simeq 0$, where $g$, $g_5$ and $g_6$ are the perihelion
precession frequencies of asteroid, Jupiter and Saturn, respectively. Due to the resonance, the orbital
eccentricity of (5) Astraea has large oscillations and a long term trend (Fig. A1). The resonant angle,
$\sigma=\varpi+\varpi_5-2\varpi_6$, can undergo transitions between libration and circulation on long
timescales (\v{S}idlichovsk\'y \& Nesvorn\'y 1996); it is therefore not possible to obtain a unique
decomposition of $e(t)$ into a quasi-periodic signal. Indeed, if we defined the proper eccentricity
as the amplitude of the (largest) proper term, the proper eccentricity value would depend on the Fourier
window that is used in FMFT. For example, for a 0.6-Myr window the amplitude of the proper term is
$\simeq 0.24$ (Fig. A2). For a 5-Myr window, the proper term undergoes splitting into several terms
with similar amplitudes; the amplitude of the largest term is $\simeq 0.12$. The length of the Fourier
window defines how many terms appear and affects the amplitude of the largest term as well.
It is difficult, using this method, to define a unique value of $e_{\rm p}$. For reference, the method
described in Section 2.1 gives $e_{\rm p}=0.171 \pm 0.012$ for (5) Astraea. 

\acknowledgements

\begin{center}
{\bf Acknowledgments} 
\end{center} 
\vspace*{-3.mm}
The simulations were performed on the NASA Pleiades Supercomputer. We thank the NASA
NAS computing division for continued support. The work of DN was funded by the NASA
SSW program. DV acknowledges support from the grant 23-04946S of the Czech Science
Foundation. FR acknowledges support from CNPq grant 312429/2023-1. We thank an anonymous
reviewer for critical comments.

%\clearpage

\begin{table}
\label{freq}
\centering
{
\begin{tabular}{lrrr}
\hline \hline
$j$ & $g_j$ & $s_j$ \\ 
    & arcsec/s       & arcsec/s \\                     
\hline                    
1 & 5.535    & -5.624\\ 
2 & 7.437    & -7.082\\
3 & 17.357   & -18.837 \\ 
4 & 17.905   & -17.749 \\
5 &  4.257   & --     \\
6 & 28.245   & -26.348 \\ 
7 &  3.088   & -2.993 \\ 
8 &  0.671   & -0.692 \\
\hline \hline
\end{tabular}
}
\caption{The secular precession frequencies of the planetary system determined from a 10-Myr-long 
numerical integration.}
\end{table}

\begin{table}
\label{catalog}  
%\rotate{90}
\centering
{ \tiny
\begin{tabular}{rrrrrrrrrrr}
\hline \hline
$a_{\rm p}$ & $\delta a_{\rm p}$ & $e_{\rm p}$ & $\delta e_{\rm p}$ & $\sin i_{\rm p}$ & $\delta \sin i_{\rm p}$ & $g$
& $s$ & $H$ & \# of & MPC \\
au        & au                &           &                   &                &
& arcsec/s  & arcsec/s & mag & opps. & desig.\\
\hline
 2.767028 & 0.23E-04 & 0.115193 & 0.16E-03 & 0.167560 & 0.13E-04   & 54.253800  & -59.249995  & 3.340 & 123 & 00001 \\  
 2.771276 & 0.12E-03 & 0.280234 & 0.34E-03 & 0.546016 & 0.45E-04   & -1.372312  & -46.451120  & 4.120 & 121 & 00002 \\  
 2.669376 & 0.75E-05 & 0.233600 & 0.95E-05 & 0.229144 & 0.25E-05   & 43.858531  & -61.476025  & 5.170 & 114 & 00003 \\  
 2.361512 & 0.59E-07 & 0.099452 & 0.14E-03 & 0.111023 & 0.18E-03   & 36.882605  & -39.610314  & 3.220 & 110 & 00004 \\  
 2.577657 & 0.29E-04 & 0.171249 & 0.12E-01 & 0.076113 & 0.14E-02   & 52.506470  & -51.132322  & 7.000  & 87 & 00005 \\  
 2.425275 & 0.15E-05 & 0.158950 & 0.13E-03 & 0.249017 & 0.52E-03   & 31.540156  & -41.819571  & 5.610 & 103 & 00006 \\  
 2.386116 & 0.27E-04 & 0.210605 & 0.51E-03 & 0.108920 & 0.48E-03   & 38.458755  & -46.352563  & 5.640  & 90 & 00007 \\  
 2.201393 & 0.20E-04 & 0.144783 & 0.12E-03 & 0.096433 & 0.17E-03   & 32.049097  & -35.508300  & 6.610  & 94 & 00008 \\  
 2.386436 & 0.32E-05 & 0.127508 & 0.84E-04 & 0.081626 & 0.49E-04   & 38.763873  & -42.014077  & 6.320  & 86 & 00009 \\  
 3.141917 & 0.59E-04 & 0.135147 & 0.93E-04 & 0.088733 & 0.49E-05  & 128.714227  & -97.051050  & 5.640  & 99 & 00010 \\  
 2.452256 & 0.18E-06 & 0.074519 & 0.22E-04 & 0.067789 & 0.15E-04   & 40.764583  & -43.166575  & 6.730 & 105 & 00011 \\  
 2.334262 & 0.39E-04 & 0.174974 & 0.64E-04 & 0.162217 & 0.61E-04   & 34.152919  & -40.893417  & 7.310  & 82 & 00012 \\  
 2.576296 & 0.12E-05 & 0.126367 & 0.11E-03 & 0.276336 & 0.48E-04   & 35.937744  & -45.446496  & 6.930  & 74 & 00013 \\  
 2.587547 & 0.14E-05 & 0.198285 & 0.15E-04 & 0.145452 & 0.59E-04   & 48.841992  & -56.278925  & 6.560  & 84 & 00014 \\  
 2.643475 & 0.32E-05 & 0.148582 & 0.81E-04 & 0.226506 & 0.50E-04   & 42.699508  & -52.035733  & 5.410  & 84 & 00015 \\  
 2.922133 & 0.76E-05 & 0.102836 & 0.23E-03 & 0.044006 & 0.24E-04   & 76.932586  & -73.291465  & 6.210  & 93 & 00016 \\  
 2.471116 & 0.71E-05 & 0.137893 & 0.16E-04 & 0.084985 & 0.30E-04   & 43.752221  & -46.311410  & 7.940  & 96 & 00017 \\  
 2.295882 & 0.14E-03 & 0.178486 & 0.57E-04 & 0.169080 & 0.46E-04   & 32.651574  & -39.374445  & 6.340  & 85 & 00018 \\  
 2.442013 & 0.17E-06 & 0.134582 & 0.17E-04 & 0.038793 & 0.24E-04   & 41.875606  & -45.185309  & 7.500  & 96 & 00019 \\  
 2.408640 & 0.80E-05 & 0.161834 & 0.28E-04 & 0.024686 & 0.46E-04   & 40.879490  & -45.112965  & 6.540  & 92 & 00020 \\  
\hline \hline\\
\end{tabular}
}
\caption{The full proper element catalog is available for download at
\texttt{https://asteroids.on.br/appeal/} and \texttt{www.boulder.swri.edu/\~{}davidn/Proper24/}.}
\end{table}

\begin{table}
\label{tab1}  
\centering
{ \small
\begin{tabular}{rllll}
\hline \hline
Number & Name & HCM cut & number    & Notes \\
       &      & (m/s)   & of mem.   &       \\
\hline
     6 & Hebe           & 80    & 112  & compact, depleted, S-type \\
    40 & Harmonia       & 70    & 160  & extended, 40 is offset, S-type \\
    67 & Asia           & 40    & 569  & diffuse, S-type \\
   115 & Thyra          & 30    & 56  & compact, S-type \\
   126 & Velleda        & 70    & 308  & extended, close to 1394, S-type \\
   135 & Hertha         & 20    & 1473  & in Nysa, big, Dykhuis et al. (2015) \\
   345 & Tercidina      & 80    & 429  & diffuse, extended in $a$, S-type \\
  1156 & Kira           & 70    & 33  & small, S-type \\
  1217 & Maximiliana    & 20    & 28  & small, convergent \\   
  1394 & Algoa          & 50    & 27  & small, close to 126, S-type \\
  1598 & Paloque        & 40    & 117  & compact, big bodies, C-type \\
  1663 & van der Bos    & 30    & 285  & in Flora, diffuse \\
  1963 & Bezovec        & 65    & 354  & in Phocaea, diffuse, offset \\
  2110 & Moore-Sitterly & 10    & 17  & compact, convergent, S-type \\
  2328 & Robeson        & 20    & 223  & compact, C-type \\
  2653 & Principia      & 50    & 131  & near Vesta, compact, S-type \\ 
  2728 & Yatskiv        & 30    & 48  & in Polana, compact, C-type \\
  2823 & van der Laan   & 60    & 136  & near Sulamitis, offset, S-type \\
  2961 & Katsurahama    & 25    & 142  & in Flora, small \\
  3452 & Hawke          & 100   & 116  & diffuse, depleted, 317?, C-type \\
  6065 & Chesneau       & 60    & 351  & in Phocaea, diffuse, 587? \\ 
  6084 & Bascom         & 10    & 10  & in Phocaea, very small, convergent \\
  6142 & Tantawi        & 30    & 114  & in Polana, compact, Novakovi\'c \& Radovi\'c (2019) \\
  8272 & Iitatemura     & 40    & 41  & compact, S-type \\
 18429 & 1994 AO1       & 20    & 38  & compact, Novakovi\'c \& Radovi\'c (2019)\\
 41331 & 1999 XB232     & 20    & 23  & in Phocaea, very small \\
 61203 & Chleborad      & 20    & 20  & in Vesta, close to 3:1, compact \\
 70208 & 1999 RX33      & 20    & 16  & in Phocaea, very small, convergent \\
118564 & 2000 FO47      & 20    & 84  & in Nysa, compact, S-type \\
484743 & 2008 YL101     & 30    & 16  & compact \\
-- & 2012 PM61          & 15    & 35  & in Vesta, compact, only small bodies \\
\hline \hline\\
\end{tabular}
}
\caption{
31 asteroid families -- not listed in Nesvorn\'y et al. (2015) -- in the inner belt ($a<2.5$ au).
Three of these families (Hertha, Tantawi and 1994 AO1) were already reported in recent publications
(Dykhuis et al. 2015, Novakovi\'c \& Radovi\'c 2019). The HCM cutoff and number of family
members identified at this cutoff are listed in columns 3 and 4, respectively. The HCM cutoffs
reported here are approximate and will be fine tuned in future publications.
}
\end{table}

\begin{table}
  \label{tab2}
\centering
{ \scriptsize
\begin{tabular}{rllll}
\hline \hline
Number & Name & HCM cut & number  & Notes \\
       &      & (m/s)   & of mem. &       \\
\hline
    28 & Bellona	     & 30  & 958   & numerous small bodies \\
    45 & Eugenia	     & 35  & 246    & 2 satellites, small bodies, diffuse, C-type \\
   177 & Irma                & 60  & 413    & diffuse, K-type? \\
   194 & Prokne              & 50  & 176    & 194 is offset, Novakovi\'c et al. (2011) \\
   237 & Coelestina          & 40  & 104    & small, S-type \\
   240 & Vanadis             & 60  & 17    & near Misa, small, C-type, needs confirmation\\
   342 & Endymion	     & 20  & 33    & very small, close to Konig, C-type \\
   351 & Irsa		     & 20  & 242    & small bodies, S-type \\
   389 & Industria           & 50  & 464    & big, S-type, offset \\
   446 & Aeternitas          & 40  & 447    & big, compact \\
   539 & Pamina              & 37  & 81    & diffuse, 539 is offset, C-type \\
   593 & Titania	     & 40  & 297    & small, C-type \\
   660 & Crescentia          & 20  & 805    & in Maria, big \\
   727 & Nipponia            & 20  & 510    & near Maria, compact, C-type \\
   801 & Helwerthia	     & 40  & 175 	& big bodies, compact, C-type \\
  1048 & Fedosia	     & 40  & 125 	& compact, C-type \\
  1127 & Mimi                & 60  & 82    & small, 1127 is offset, C-type \\
  1160 & Illyria             & 30  & 692    & in Maria, diffuse \\
  1347 & Patria              & 50  & 132    & compact, 1347 is offset, C-type \\
  2079 & Jacchia             & 35  & 125    & in Eunomia, diffuse \\
  2927 & Alamosa	     & 20  & 14 	& very small \\
  3324 & Avsyuk		     & 30  & 222 	& offset, S-type \\
  3497 & Innanen	     & 50  & 261    & small, C-type \\
  3567 & Alvema              & 55  & 260    & extended, big bodies, C-type \\
  5798 & Burnett             & 58  & 81    & extended, diffuse, S-type \\
  7233 & Majella             & 10  & 17    & very small \\
  7403 & Choustnik           & 30  & 57    & small, 7403 is offset, C-type \\
  8223 & Bradshaw            & 20  & 48    & very small, C-type \\
  9332 & 1990 SB1            & 10  & 8   & in Eunomia, compact, convergent \\
 10164 & Akusekijima         & 10  & 18    & compact, convergent\\
 11014 & Svatopluk           & 50  & 159    & extended, small bodies, C-type \\
 12586 & Shukla		     & 20  & 25    & very small, S-type \\
 15104 & 2000 BV3	     & 30  & 78 	& small, C-type \\
 16472 & 1990 OE5            & 7 &  58   & in Dora, compact, C-type \\
 16940 & 1998 GC3            & 20  & 40    & in Eunomia, diffuse \\
 21591 & 1998 TA6            & 15  & 47    & in Eunomia, compact \\
 22766 & 1999 AE7	     & 10  & 5    & compact, needs confirmation \\
 26170 & Kazuhiko	     & 10  & 9 	& compact, S-type \\
 30718 & Records             & 40  & 95    & diagonal, C-type \\
 32983 & 1996 WU2            & 20  & 50    & in Eunomia, small \\
 42357 & 2002 CS52           & 40  & 216    & compact, S-type \\
 49362 & 1998 WW16	     & 30  & 68 	& 49362 is offset, S-type \\
 53209 & 1999 CQ75	     & 20  & 189 	& small, C-type \\
 54934 & 2001 OH105	     & 50  & 37 	& small bodies \\
190237 & 2007 DM10	     & 50  & 41 	& small \\
249576 & 1995 FH2	     & 30  & 78 	& diffuse, C-type \\
435544 & 2008 OV23           & 20  & 90    & in Eunomia, compact, small bodies \\
--     & 2006 QX49	     & 20  & 11 	& compact \\
\hline \hline\\
\end{tabular}
}
\caption{
48 asteroid families -- not listed in Nesvorn\'y et al. (2015) -- in the middle belt ($2.5<a<2.82$ au).
The Prokne family has already been reported in Novakovi\'c et al. (2011).
The HCM cutoff and number of family
members identified at this cutoff are listed in columns 3 and 4, respectively. The HCM cutoffs
reported here are approximate and will be fine tuned in future publications.
}
\end{table}

\begin{table}
  \label{tab3}
  \centering
  {
\begin{tabular}{rllll}
  \hline \hline
Number & Name & HCM cut & number  & Notes \\
       &      & (m/s)   & of mem. &       \\
\hline
   174 & Phaedra             & 60 & 170 & diffuse, S-type \\
   321 & Florentina          & 10 & 209  & Koronis$_4$ in Bro\v{z} et al. (2024), HCM cut \\
   392 & Wilhelmina          & 40 & 45  & compact, C-type \\
   924 & Toni                & 50 & 50  & compact, C-type \\
  1289 & Kutaisii            & 10 & 371  & Koronis$_3$ in Bro\v{z} et al. (2024) \\
  3787 & Aivazovskij         & 15 & 12  & in Itha, compact \\
  4471 & Graculus            & 50 & 136  & small, S-type \\
%  6457 & Kremsmunster           & in Koronis \\ this one is hard to identify, seen in e/i projection
 11048 & 1990 QZ5            & 50 & 35  & extended in $e$, compact in $i$ \\
 15454 & 1998 YB3            & 50 & 340  & large bodies, C-type \\
 19093 & 1979 MM3            & 90 & 71  & split of 709 Fringilla (FIN 623) \\
 26369 & 1999 CG62           & 50 & 43  & small, C-type \\
 31810 & 1999 NR38           & 60 & 103   & diffuse, S-type \\
 37981 & 1998 HD130          & 90 & 419  & split of 709 Fringilla (FIN 623) \\ 
 77873 & 2001 SQ46           & 50 & 21  & small, C-type \\
 78225 & 2002 OS10           & 60 & 88  & diffuse \\
211772 & 2004 BQ90           & 70 & 295  & diffuse, C-type \\
217472 & 2005 WV105          & 40 & 37  & small \\
\hline \hline\\
\end{tabular}
}
\caption{
17 asteroid families -- not listed in Nesvorn\'y et al. (2015) -- in the pristine zone ($2.82<a<2.96$ au).
The Florentina and Kutaisii families were reported in Bro\v{z} et al. (2024) as Koronis$_4$ and Koronis$_3$,
respectively. The (19093) 1979MM3 and (37981) 1998HD130 families are a split of the (709) Fringilla family (FIN 623;
Nesvorn\'y et al. 2015) into two parts.
The HCM cutoff and number of family
members identified at this cutoff are listed in columns 3 and 4, respectively. The HCM cutoffs
reported here are approximate and will be fine tuned in future publications.
%For the Florentina family, HCM needs to be restricted to bodies with $\sin i > 0.03785$.
}
\end{table}

\begin{table}
  \label{tab4}
\centering
{ \footnotesize
\begin{tabular}{rllll}
  \hline \hline
Number & Name & HCM cut & number    & Notes \\
       &      & (m/s)   & of mem.   &       \\
\hline
    52 & Europa               & 50 & 250 & diffuse, C-type \\
   106 & Dione                & 55 & 261 & extended, diffuse, 3136? \\
   260 & Huberta	      & 70 & 190 & extended, C-type, Cybeles, Carruba et al. (2015)\\
   286 & Iclea		      & 50 & 143	& compact, C-type \\
   414 & Liriope              & 70 & 14 & compact, depleted, Cybeles \\
   633 & Zelima               & 10 & 88 & in Eos, compact, z$_1$ resonance, Tsirvoulis (2019) \\
   690 & Wratislavia	      & 30 & 820	& big, C-type \\
   850 & Altona               & 60 & 73 & depleted \\
   885 & Ulrike		      & 30 & 45 & compact, C-type \\
   991 & McDonalda            & 20 & 260 & in Themis, diffuse \\
  1323 & Tugela               & 50 & 412 & big, C-type, 1323 offset, 69559 in Novakovi\'c et al. (2011)\\
  1357 & Khama		      & 30 & 62 & compact, C-type \\
  1461 & Jean-Jacques	      & 30 & 174	& compact, numerous small bodies, M-type? \\
  1524 & Joensuu              & 40 & 201 & K-type? \\
  1599 & Giomus               & 25 & 322 & in Hygiea \\
  2458 & Veniakaverin	      & 30 & 177	& in Themis \\
  2562 & Chaliapin            & 22 & 271 & in Eos, diffuse \\
  3310 & Patsy                & 15 & 826 & in Eos, compact \\
%  3600 & Archimedes             & BIG, S-type, former Rafita, FIN 518 \\
  3803 & Tuchkova	      & 10 & 7	& small, C-type \\
  4291 & Kodaihasu            & 20 & 16 & compact \\
  4897 & Tomhamilton	      & 20 & 254	& in Eos, 4897 is offset, diffuse \\
  5228 & Maca		      & 30 & 47	& in Themis, very small \\
  6924 & Fukui		      & 80 & 146	& extended, C-type, Cybeles, Carruba et al. (2019) \\
  7504 & Kawakita             & 10 & 19 & small, C-type \\
  8737 & Takehiro             & 40 & 569 & 8737 is offset, HCM cut \\
  9522 & Schlichting          & 100 & 9 & only a few big bodies, Cybeles, Carruba et al. (2019)\\
 12911 & Goodhue              & 18 & 132 & in Themis \\
 20674 & 1999 VT1	      & 10 & 35	& compact, 20674 offset, Gibbs in Novakovi\'c et al. (2014) \\
 29880 & Andytran	      & 20 & 44  & in Themis \\
 34216 & 2000 QK75	      & 20 & 22	& small, C-type \\
 37455 & 4727 P-L	      & 40 & 285	& 37455 is offset, C-type \\
 48412 & 1986 QN1             & 60 & 100 & compact \\
 48506 & 1993 FO10            & 20 & 90 & near Tirela, C-type \\
 63235 & 2001 BV20            & 60 & 35 & small, C-type, Cybeles \\
 77882 & 2001 SV124           & 45 & 316  & diffuse, HCM cuts, needs confirmation\\
106302 & 2000 UJ87	      & 30 & 489	& extended, C-type, 86 is offset, 13:6 diffusion?\\
157940 & 1999 XU240	      & 10 & 19	& compact, low number of bodies \\
352479 & 2008 BO31	      & 30 & 50	& small \\
365736 & 2010 WK8	      & 50 & 40	& small, C-type, Cybeles \\
\hline \hline\\
\end{tabular}
}
\caption{
39 asteroid families -- not listed in Nesvorn\'y et al. (2015) -- in the outer belt ($a>2.96$ au).
Five of these families (Huberta, Zelima, Tugela, Fukui, Schlichting, Gibbs/1999 VT1) were already reported
in previous publications (Novakovi\'c et al. 2011, 2014; Carruba et al. 2015, 2019; Tsirvoulis 2019).
The HCM cutoff and number of family
members identified at this cutoff are listed in columns 3 and 4, respectively. The HCM cutoffs
reported here are approximate and will be fine tuned in future publications.
}
\end{table}

\begin{table}
  \label{tab5}
  { \small
\centering
\begin{tabular}{rllll}
\hline \hline
Number & Name & HCM cut & number & Notes \\
       &      & (m/s)   & of mem.          &       \\
\hline
   183 & Istria               & 80  & 301 & big, S-type \\
   350 & Ornamenta            & 70 & 716 & in Alauda, big, extended, C-type, 130? \\
   386 & Siegena              & 100 & 168  & extended, C-type \\
   522 & Helga                & 200 & 34 & extended, C-type, Cybeles, Carruba et al. (2015)\\
   704 & Interamnia           & 50 & 413 & offset, cratering?, C-type \\
   754 & Malabar	      & 60 & 79	& extended, also big bodies, C-type \\
  1312 & Vassar		      & 30 & 21	& C-type, Novakovi\'c et al. (2011)\\
  1390 & Abastumani	      & 40 & 65	& small, C-type, Cybeles, Carruba et al. (2019)\\
  3001 & Michelangelo         & 200 & 42 & dispersed, no background \\
  3854 & George               & 130 & 261 & in Hungarias, extended, diffuse, S-type \\
  7605 & Cindygraber	      & 30 & 85 & small, C-type, Novakovi\'c et al. (2011)\\
 63530 & 2001 PG20            & 90 & 419 & extended, C-type \\
 77899 & 2001 TS117           & 200 & 257 & very extended, low-density, 77899 is offset, M-type? \\
 78705 & 2002 TE180	      & 10 & 24	& compact, S-type \\
101567 & 1999 AW22	      & 50 & 293	& extended, C-type? 116763 in Novakovi\'c et al. (2011)\\
126948 & 2002 FX3	      & 30 & 14	& compact, low number of bodies, C-type \\
236657 & 2006 KU114	      & 70 & 303	& extended, C-type \\
316974 & 2001 FP147           & 150 & 53 & dispersed, no background, C-type \\
\hline \hline\\
\end{tabular}
}
\caption{
18 asteroid families -- not listed in Nesvorn\'y et al. (2015) -- with high orbital inclinations
  ($\sin i > 0.3$).
Five of these families (Helga, Vassar, Abastumani, Cindygraber, 116763/1999 AW22) were already reported in
previous publications (Novakovi\'c et al. 2011, Carruba et al. 2019). The HCM cutoff and number of family
members identified at this cutoff are listed in columns 3 and 4, respectively. The HCM cutoffs
reported here are approximate and will be fine tuned in future publications.
}
\end{table}

\begin{table}
  \label{tab8}
\centering
\begin{tabular}{rll}
\hline \hline
Number & Name & Probability\\
\hline
106 & Dione                           & $1\times10^{-6}$ \\   
240 & Vanadis                         & 0.003           \\
414 & Liriope                         & $2\times10^{-6}$ \\    
2110 & Moore-Sitterly                 & $2\times10^{-7}$ \\
2927 & Alamosa                        & $3\times10^{-7}$ \\
3787 & Aivazovskij                    & $6\times10^{-6}$ \\
3803 & Tuchkova                       & $6\times10^{-4}$ \\
6084 & Bascom                         & $5\times10^{-5}$ \\
7233 & Majella                        & $2\times10^{-7}$ \\
9332 & 1990SB1                        & 0.02            \\
9522 & Schlichting                    & $5\times10^{-5}$ \\
22766 & 1999AE7                       & 0.01            \\
26170 & Kazuhiko                      & $3\times10^{-4}$ \\
77882 & 2001SV124                     & 0.05            \\ 
126948 & 2002FX3                      & $3\times10^{-7}$ \\

--    & 2006 QX49                     & $7\times10^{-6}$ \\ 

\hline \hline\\
\end{tabular}
\caption{
  Asteroid families for which we were able to distinguish the statistical significance from 1 (from
  $10^7$ trials with the box method; see Sect. 2.2). The third column reports the probability that
  random fluctuations of bodies in the local background could produce the family in question.
}
\end{table}

\begin{table}
\label{moore}
  \centering
{ \footnotesize
\begin{tabular}{rrrrrrrr}
\hline \hline
 MPC     &    $H$  &      $a$   &      $e$  &    $i$  &    $\Omega$ & $\omega$ &  $M$ \\
 desig.  &    (mag)  &   (au)   &           &    (deg)&    (deg) &    (deg) &   (deg) \\
\hline
 02110   & 13.63    & 2.1980514 & 0.1770290 & 1.13155 & 140.45313 & 192.69746 & 338.26094 \\  
 44612   & 15.67    & 2.1976575 & 0.1779052 & 1.12194 & 141.53740 & 189.48402 & 207.21536 \\  
 X8073   & 18.66    & 2.1965096 & 0.1762623 & 0.88896 & 159.45456 & 160.96067 & 230.71251 \\  
 q9915   & 18.75    & 2.1974504 & 0.1763956 & 0.91317 & 157.56482 & 163.77923  & 87.46958 \\  
 K01K82Y & 19.10  & 2.1969377 & 0.1759087 & 0.93567 & 155.79653 & 166.29243 & 347.34205 \\  
 K02F44D & 18.79  & 2.1965174 & 0.1770381 & 0.92887 & 155.96118 & 166.21130 & 135.31503 \\  
 K06UN8L & 19.35  & 2.1958976 & 0.1755210 & 0.83123 & 165.46635 & 151.66292 & 182.21323 \\  
 K10RG7V & 19.54  & 2.1970673 & 0.1740991 & 0.78330 & 170.10353 & 145.11975  & 82.11909 \\  
 K13V79C & 19.13  & 2.1961976 & 0.1770934 & 0.93426 & 156.16395 & 165.77682 & 133.44064 \\  
 K14Qw7S & 19.75 & 2.1959868 & 0.1753686 & 0.88033 & 161.03410 & 158.30843 & 356.31624  \\ 
 K15RJ4A & 19.17   & 2.1964264 & 0.1750963 & 0.82816 & 165.61774 & 151.71076 & 258.86629 \\  
 K15T83O & 19.20  & 2.1968660 & 0.1762778 & 0.92017 & 156.87362 & 164.62329 & 273.37375  \\ 
 K16S14Q & 19.29  & 2.1961233 & 0.1768452 & 0.91131 & 157.83567 & 163.15490 & 140.52663  \\ 
 K19O06U & 19.90  & 2.1955252 & 0.1751064 & 0.79545 & 168.73592 & 147.83085 & 158.42254  \\ 
 K19SB1L & 19.48   & 2.1956555 & 0.1752386 & 0.79853 & 168.11064 & 147.76117 & 185.50486 \\  
 K21N62V & 19.93  & 2.1961285 & 0.1740565 & 0.78380 & 170.20425 & 144.75362 & 292.68934 \\  
 K22O48Q & 19.43  & 2.1956928 & 0.1752722 & 0.81904 & 166.57575 & 149.94720 & 196.72505  \\ 
\hline \hline\\
\end{tabular}
}
\caption{The osculating orbits of 17 members of the Moore-Sitterly family (MJD epoch 60400.0). 
Note the clustering of nodal ($\Omega$) and perihelion ($\omega$) longitudes.}
\end{table}

\begin{table}
\label{9332}
  \centering
{ \footnotesize
\begin{tabular}{rrrrrrrr}
\hline \hline
 MPC     &    $H$  &      $a$   &      $e$  &    $i$  &    $\Omega$ & $\omega$ &  $M$ \\
 desig.  &    (mag)  &   (au)   &           &    (deg)&    (deg) &    (deg) &   (deg) \\
\hline
09332   & 13.24 & 2.5826279 & 0.1071552 & 12.64528 & 346.35641 & 211.88757 & 200.68128 \\
$\sim$0EdI   & 18.11 & 2.5791555 & 0.1091825 & 12.69705 & 346.37901 & 210.86424 & 322.66213\\ 
K08CA6R & 17.76 & 2.5795956 & 0.1090338 & 12.67301 & 346.36232 & 211.83755 & 277.44882 \\
K10D44Y & 17.74 & 2.5808240 & 0.1057713 & 12.65371 & 346.31579 & 211.24047 & 120.44629\\ 
K15Ra4C & 18.02 & 2.5811666 & 0.1068327 & 12.64488 & 346.29494 & 211.84784 & 167.30554\\ 
K16D45B & 17.99 & 2.5789990 & 0.1093170 & 12.69066 & 346.38013 & 211.21348 & 307.40249\\ 
K16E06Q & 18.67 & 2.5790002 & 0.1093096 & 12.69115 & 346.38105 & 211.19633 & 308.01388\\ 
K17E36K & 18.16 & 2.5826844 & 0.1076307 & 12.64840 & 346.34149 & 212.00629 & 234.08503 \\
K17FJ3L & 17.79 & 2.5817726 & 0.1080386 & 12.65332 & 346.33790 & 212.05776 & 245.77392 \\
\hline \hline\\
\end{tabular}
}
\caption{The osculating orbits of 9 members of the (9332) 1990SB1 family (MJD epoch 60400.0). 
Note the clustering of nodal ($\Omega$) and perihelion ($\omega$) longitudes.}
\end{table}

\clearpage

\clearpage

\clearpage

%\clearpage
%\begin{figure}
%\epsscale{0.8}
%\plotone{cat2.eps}
%\caption{}
%\label{cat2}
%\end{figure}

%\clearpage
%\begin{figure}
%\epsscale{0.8}
%\plotone{cat2b.eps}
%\caption{}
%\label{cat2b}
%\end{figure}

\clearpage

\clearpage

\clearpage

\clearpage

\clearpage

\clearpage

\clearpage

\clearpage

\clearpage

\clearpage
\begin{figure}
\epsscale{0.8}
\plotone{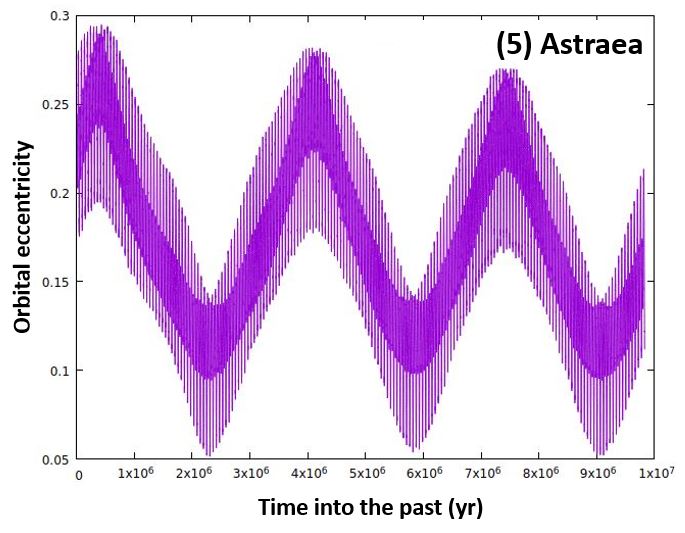}
\caption{{\bf Fig A1.} The orbital eccentricity of (5) Astraea shows large oscillations due to the
interaction with the secular resonance $g+g_5-2g_6$.}
\label{figa1}
\end{figure}

\clearpage
\begin{figure}
\epsscale{0.8}
\plotone{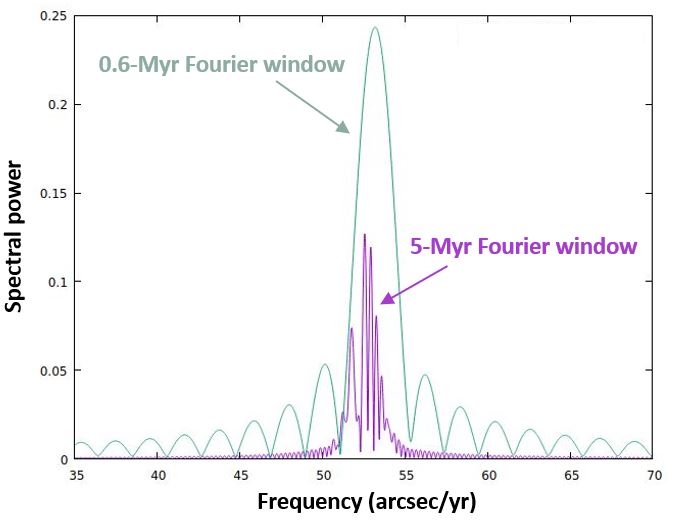}
\caption{{\bf Fig. A2} The Fourier spectrum of $e\cos(\varpi),e\sin (\varpi)$ obtained for (5) Astraea
  from two different intervals: 0.6 Myr in green and 5 Myr in purple. The Fourier spectrum
  obtained from the 5-Myr long interval illustrates splitting of the proper frequency into
  several terms.}
\label{figa2}
\end{figure}

\end{document}